\documentclass[prd,superscriptaddress,nofootinbib,longbibliography]{revtex4-1}
\usepackage{graphicx}
\usepackage{amsmath}
\usepackage{amsfonts}
\usepackage{mathtools}
\usepackage{color}
\usepackage{amssymb}
\usepackage{dcolumn}% Align table columns on decimal point
\usepackage{bm}% bold math
\usepackage{braket}
\usepackage{microtype} %makes everything nicer
\usepackage[linktoc=all]{hyperref}
\usepackage{yfonts}

\newcommand{\lag}{\mathcal{L}}
\newcommand{\ud}{\mathrm{d}}

\newenvironment{itemize*}
  {\begin{itemize}
    \setlength{\itemsep}{0pt}
    \setlength{\parskip}{0pt}}
  {\end{itemize}}

\newenvironment{enumerate*}
  {\begin{enumerate}
    \setlength{\itemsep}{0pt}
    \setlength{\parskip}{0pt}}
  {\end{enumerate}}

\newenvironment{description*}
  {\begin{description}
    \setlength{\itemsep}{0pt}
    \setlength{\parskip}{0pt}}
  {\end{description}}

\newenvironment{eqn}
	{
		\equation 
		\aligned
	}	
	{
		\endaligned
		\endequation
	}
\def\ben{\begin{enumerate*}}
\def\een{\end{enumerate*}}
\def\bi{\begin{itemize*}}
\def\ei{\end{itemize*}}
\def\bd{\begin{description*}}
\def\ed{\end{description*}}
\def\be{\begin{equation}}
\def\ee{\end{equation}}
\def\bea{\begin{eqnarray}}
\def\eea{\end{eqnarray}}
\def\bfl{\begin{flushleft}}
\def\efl{\end{flushleft}}
\def\bigskip{\;\;\;\;\;\;\;}

\newcommand{\bs}[1]{{\boldsymbol{#1}}}

\begin{document}

\begin{flushright}
QMUL-PH-19-11
\end{flushright}

\title{The classical double copy in three spacetime dimensions}

\author{Mariana Carrillo Gonz\'alez}
\email{cmariana@sas.upenn.edu}
\affiliation{Center for Particle Cosmology, Department of Physics and Astronomy,
University of Pennsylvania, Philadelphia, Pennsylvania 19104, USA}

\author{Brandon Melcher} \email{bsmelche@syr.edu} \author{Kenneth
  Ratliff} \email{krratlif@syr.edu} \author{Scott Watson}
\email{gswatson@syr.edu} \affiliation{Department of Physics, Syracuse
  University, Syracuse, NY 13244, USA}

\author{Chris D. White} \email{christopher.white@qmul.ac.uk}
\affiliation{Centre for Research in String Theory, School of Physics
  and \\ Astronomy, Queen Mary University of London, 327 Mile End Road,
  London E1 4NS, UK}

\date{\today}

\begin{abstract}
The double copy relates scattering amplitudes in gauge and gravity
theories, and has also been extended to classical solutions. In this
paper, we study solutions in three spacetime dimensions, where the
double copy may be expected to be problematic due to the absence of
propagating degrees of freedom for the graviton, and the lack of a
Newtonian limit. In particular, we examine the double copy of a gauge
theory point charge. This is a vacuum solution in gauge theory, but
leads to a non-vacuum solution in gravity, which we show is consistent
with previously derived constraints. Furthermore, we
successfully interpret the non-trivial stress-energy tensor on the
gravity side as arising from a dilaton profile, and the Newtonian
description of a point charge emerges as expected in the appropriate
limit. Thus, our results provide a non-trivial cross-check of
the classical Kerr-Schild double copy.
\end{abstract}

%\pacs{}
\maketitle
\thispagestyle{empty}
%%%%%%%%%%%%%%%%%%%%%%%%%%%%%%% 
\newpage

\section{Introduction}
\label{sec:intro}

The BCJ double copy, originally proposed in \cite{Bern:2008qj},
relates scattering amplitudes in gauge and gravity
theories. Specifically, it invokes {\it color-kinematic duality},
which can be defined as follows. First, one may express an $m$-point
$\ell$-loop gluon amplitude as a sum over cubic graphs:
		\begin{eqn}
			\mathcal{A}^\ell_m = \sum_g \frac{c(g)
                          n(g)}{d(g)}
		\end{eqn}
where $c(g)$, $n(g)$, and $d(g)$ are the color factors, kinematic numerators,
and propagators respectively of the cubic graph $g$, and integrals
over loop momenta are left implicit. Individual terms in the sum are
not unique, but mix with each other under {\it generalized gauge
  transformations}, consisting of conventional (local) gauge
transformations and / or field redefinitions. Color-kinematics duality
is then the statement that there exists a choice of generalized gauge,
called color-dual gauge, in which the kinematic numerators obey the
same algebra as the color factors:
		\begin{eqn}
			c(g)
				&= - c(\bar{g}),
		&
			n(g)
				&= n(\bar{g}),
		\\
			c(g_s)
				&= c(g_t) + c(g_u),
		&
			n(g_s)
				&= n(g_t) + n(g_u),
		\end{eqn}
where $\bar{g}$ denotes a graph obtained from $g$ by exchanging an odd
number of vertices and the graphs $g_t$ and $g_u$ are obtained from
the graph $g_s$ by picking an internal leg in $g_s$, calling it the
$s$-channel, and then switching it to the $t$- or $u$-channel, leaving
the rest of the graph unchanged. The relations obeyed by the $c$'s are
automatic, following solely from the algebraic properties of the
structure constants of Lie groups (i.e. the Jacobi identity). On the other hand, the fact that we can find kinematic numerators satisfying the same  algebra is highly non-trivial and does not hold  for any gauge  invariant operator \cite{Broedel:2012rc,He:2016iqi,Cofano:2015jva}. The BCJ
double copy then states that given some such $\mathcal{A}^\ell_m$ in
color-dual gauge we can obtain an $m$-point $\ell$-loop graviton
amplitude by replacing the color factors with another set of kinematic
numerators,
		\begin{eqn}
			\mathcal{M}^\ell_m
				= \sum_g \frac{n(g) \tilde{n}(g)}{d(g)},
		\end{eqn}
where the $\tilde{n}$'s are also in color-dual gauge and need not come
from the same gauge theory. The two contributing gauge theories are
called the {\it single copies} of the gravity theory, which is called
the {\it double copy} of the two gauge theories, where the same
nomenclature is applied to the amplitudes themselves. The double copy
(and / or BCJ duality) is proven at tree
level~\cite{BjerrumBohr:2009rd,Stieberger:2009hq,Bern:2010yg,BjerrumBohr:2010zs,Feng:2010my,Tye:2010dd,Mafra:2011kj,Monteiro:2011pc,BjerrumBohr:2012mg},
where it is known to be equivalent to the so-called KLT relations in
string theory~\cite{Kawai:1985xq}. There is also highly non-trivial
evidence at loop
level~\cite{Bern:2010ue,Bern:1998ug,Green:1982sw,Bern:1997nh,Carrasco:2011mn,Carrasco:2012ca,Mafra:2012kh,Boels:2013bi,Bjerrum-Bohr:2013iza,Bern:2013yya,Bern:2013qca,Nohle:2013bfa,
  Bern:2013uka,Naculich:2013xa,Du:2014uua,Mafra:2014gja,Bern:2014sna,
  Mafra:2015mja,He:2015wgf,Bern:2015ooa,
  Mogull:2015adi,Chiodaroli:2015rdg,Bern:2017ucb,Johansson:2015oia,Bern:2010ue,Carrasco:2015iwa},
to all-orders in certain kinematic
limits~\cite{Oxburgh:2012zr,White:2011yy,Melville:2013qca,Luna:2016idw,Saotome:2012vy,Vera:2012ds,Johansson:2013nsa,Johansson:2013aca,Luna:2016idw},
and including in the partial or complete absence of supersymmetry. In
particular, the double copy of pure non-supersymmetric Yang-Mills
theory is Einsteinian gravity coupled to a two-form field and a
dilaton.

Both the color-kinematic duality and BCJ double copy described above
are purely perturbative relations, but their seeming validity at
multiple loops indicates that they may be the perturbative
manifestations of some nonperturbative symmetry or duality between
gauge and gravity theories, though it is as yet unknown how they may
be made explicit at the Lagrangian level (see
e.g. refs.~\cite{Bern:1999ji,Cheung:2016say,Cheung:2017kzx,Cheung:2017ems}
for an exploration of the latter). Motivated by this possibility,
ref.~\cite{Monteiro:2014cda} extended the double copy from scattering
amplitudes to a certain family of exact classical
solutions. Specifically, it considered stationary Kerr-Schild metrics,
such that there exists a coordinate system in which its components
satisfy
		\begin{eqn}
			g_{\mu \nu}
				&= \eta_{\mu \nu} + \kappa \phi k_\mu k_\nu,
		&
			\partial_0 g_{\mu \nu}
				= 0,
                                \label{KSdef}
		\end{eqn}
with $\eta_{\mu \nu}$ the Minkowski metric, $k$ a vector field which
is null and geodesic with respect to $\eta$ and $g$, and $\phi$ a
scalar field. Reference~\cite{Monteiro:2014cda} then showed that the
Einstein equations for $g_{\mu \nu}$ imply that the vector field
$A_\mu = \phi k_\mu$ satisfies the abelian Yang-Mills equations of
motion 
		\begin{equation}
			\partial_\nu F^{\mu \nu} 
				= g J^\mu,
		\label{Maxwell}
		\end{equation}
and is thus a well-defined single copy of the graviton
$h_{\mu\nu}$. Subsequent studies extended this class of solutions to
include multiple Kerr-Schild terms~\cite{Luna:2015paa}, and time
dependence~\cite{Luna:2016due}, as well as examining how source terms
are related in the two theories~\cite{Ridgway:2015fdl}. Furthermore,
it has been shown that classical solutions can be related through the double
copy even if they cannot be put in Kerr-Schild form at the expense of having to work order-by-order in
perturbation
theory~\cite{Neill:2013wsa,Luna:2016hge,Luna:2017dtq,Goldberger:2016iau,Goldberger:2017frp,Goldberger:2017vcg,Goldberger:2017ogt,Bern:2019nnu,Chester:2017vcz,Shen:2018ebu,Plefka:2018dpa}. Investigations
of non-perturbative aspects have been carried out in
refs.~\cite{Berman:2018hwd,White:2016jzc,DeSmet:2017rve,Bahjat-Abbas:2018vgo},
and an alternative body work has looked at matching up gauge and
gravity solutions in a wide catalogue of linearised theories, in
arbitrary
gauges~\cite{Anastasiou:2014qba,Borsten:2015pla,Anastasiou:2016csv,Anastasiou:2017nsz,Anastasiou:2017taf,Anastasiou:2018rdx,Nagy:2014jza,Cardoso:2016ngt,LopesCardoso:2018xes}. In
refs.~\cite{Carrillo-Gonzalez:2017iyj,Bahjat-Abbas:2017htu}, an
extension of the Kerr-Schild approach to non-Minkowski background
metrics was investigated, where an interesting byproduct of
ref.~\cite{Carrillo-Gonzalez:2017iyj} is a general relation that must
be satisfied by the source current in the gauge theory, and the
stress-energy tensor in the gravity theory.  For stationary solutions this relation  reads\footnote{The
  non-covariant-looking presence of the components ${T^\mu}_0$ comes
  from the also-non-covariant assumption that there exists a
  particular timelike direction along which $g_{\mu \nu}$ is
  constant. A covariant form which involves a general Killing vector field can be found in ref.~\cite{Carrillo-Gonzalez:2017iyj}.}:
		\begin{eqn}\label{eq:single_copy_source}
			J^\mu
				&= 2 \bigg( {\delta^\mu}_0 \frac{T}{d-2} -
					{T^\mu}_0  \bigg),
		\end{eqn}
	where $T = {T^{\mu}}_\mu$. For completeness we note that Eq. (\ref{eq:single_copy_source}) holds only if we choose $k^0 = +1$ and hence differs by a minus sign 
from the corresponding expression in \cite{Carrillo-Gonzalez:2017iyj}, 
in which $k_0 = +1$.

Despite the above developments, a full understanding of the double
copy, including a possible underlying explanation, is still
lacking. It is then interesting to consider highly unusual situations,
in which the double copy might break. If it survives such non-trivial
tests, our confidence in the ultimate validity of the double copy is
increased, and the frontiers of its potential application extended. A
suitable playground in this regard is to consider three spacetime
dimensions. As we review in what follows, General Relativity in $d=3$
is straightforward to formulate, but has the unfortunate consequence
that there are no propagating degrees of freedom for the graviton, and
no well-behaved Newtonian limit. This is not a problem for gauge
theory in three spacetime dimensions, which from the point of view of
the Kerr-Schild single copy means the linearised Yang-Mills, or
Maxwell, equations. In the latter case, one can easily speak of a
propagating photon, and thus one might worry that degrees of freedom
no longer match up in the Kerr-Schild double copy, which typically
associates a gauge theory solution with a pure gravity counterpart, as
above.

The ultimate resolution of this puzzle is that the double copy of
gauge theory is not pure gravity but, as noted above, GR coupled to a
dilaton and two-form field. This provides a possible mechanism for the
Kerr-Schild double copy to work, but if and how it works must be
investigated using explicit applications. We will present two examples
in this paper. The first is the well-known BTZ black
hole~\cite{Banados:1992wn}, first studied in the context of the
Kerr-Schild single copy in ref.~\cite{Carrillo-Gonzalez:2017iyj}. This
is an example without a non-zero dilaton profile on the gravity
side. Our second example is entirely new and consists of the double
copy of a point charge in (Abelian) gauge theory. Although the point
charge is a vacuum solution in the gauge theory (apart from at the
origin, where the charge resides), we will see that its double copy is
not a vacuum solution of the Einstein's equations. Instead, there is a
non-zero stress-energy tensor throughout spacetime, which will be
successfully interpreted as arising from a dilaton. Thus, the double
copy relation indeed survives and provides a highly non-trivial
example of the Kerr-Schild double copy: in no previous examples has
the dilaton been non-zero. We will see a nice physical interpretation
of its presence here, namely that it leads to geodesics that can
follow stable orbits and have a well-defined Newtonian-like limit
mimicking the behaviour in the gauge theory.

The structure of our paper is as follows. In section~\ref{sec:2+1}, we
review the issues concerning gravity in 2+1 dimensions, before
examining the Kerr-Schild copy of the previously studied BTZ black
hole in section~\ref{sec:BTZ}. In section~\ref{sec:pointcharge} we
construct the double copy of the point charge in 2+1 dimensions, and
examine its physical properties. We discuss our results and conclude
in section~\ref{sec:conclude}.

\subsection*{Conventions}  

We denote the gravitational coupling by $\kappa$ and define it in arbitrary dimensions by the Einstein-Hilbert action, or equivalently the Einstein equations:
		\begin{eqn}
			S_\text{EH}
				&= \int \text{d}^d x \, \sqrt{-g} \bigg( \frac{1}{\kappa^2} R 
					+ \mathcal{L}_\text{matter} \bigg)
		\iff
			G_{\mu \nu}
				&= \frac{\kappa^2}{2} T_{\mu \nu}.
		\end{eqn}
We define Newton's constant $G_d$ in $d$ dimensions to be the constant of proportionality between the Newtonian potential of a unit mass and its $r$-dependence. In four spacetime dimensions these definitions implies that $\kappa$ is related to Newton's constant $G_4$ and the (reduced) Planck mass $M_\text{Pl}$ by
		\begin{eqn}
			\kappa^2
				= 16 \pi G_4
				= \frac{2}{M_\text{Pl}^2}.
		\end{eqn}
However in other dimensionalities this relationship will change.

We denote a generic gauge coupling constant by $g$. When relating classical double and single copies we substitute coupling constants as
		\begin{eqn}
			\frac{\kappa}{2}
				\leftrightarrow g.
		\end{eqn}

In Kerr-Schild contexts we raise and lower all indices with the Minkowski metric $\eta_{\mu \nu}$, with the exception of the indices on $R_{\mu \nu}$ and $T_{\mu \nu}$ which we raise and lower with the full metric $g_{\mu \nu}$.

We use the mostly plus metric signature $(-,+,+,+)$.

\section{The Classical Double Copy}
\label{sec:cdc_review}
The classical single copy of a stationary $d$-dimensional Kerr-Schild spacetime
		\begin{eqn}
			g_{\mu \nu}
				&= \eta_{\mu \nu} + \kappa k_\mu k_\nu \phi.
		\end{eqn}
	can be identified as follows. The Ricci tensor with mixed indices is given by \cite{Stephani:2003tm}
		\begin{eqn}
			{R^\mu}_\nu
				&= \frac{\kappa}{2} \bigg\{
						\partial^\lambda \partial^\mu (\phi k_\lambda k_\nu)
						+ \partial^\lambda \partial_\nu (\phi k^\mu k_\lambda)
						- \partial^2 (\phi k^\mu k_\nu)
					\bigg\},
		\\
				&= \frac{\kappa^2}{2} \left\{ {T^\mu}_\nu
					- \frac{1}{d-2} {\delta^\mu}_\nu T
					\right\},
		\end{eqn}
	in the last step citing the trace-reversed Einstein equations (with $T = {T^\mu}_\mu$). Choosing $k^0 = \pm 1$ and defining $A_\mu = k_\mu \phi$, $F_{\mu \nu} = \partial_\mu A_\nu - \partial_\nu A_\mu$, the ${R^\mu}_0$ equations become
		\begin{eqn}\label{eq:single_copy_almost}
			\partial_\nu ( \partial^\mu (\phi k^\nu) - \partial^\nu (\phi k^\mu))
				&= \pm \kappa \left\{ \frac{1}{d-2} {\delta^\mu}_0 T 
					- {T^\mu}_0 \right\}.
		\end{eqn}
	Hence if we identify $A_\mu = k_\mu \phi$ and
		\begin{eqn}
			J^\mu
				&= \pm 2 \left\{ \frac{1}{d-2} {\delta^\mu}_0 T 
					- {T^\mu}_0 \right\}
		\end{eqn}
	then Eq. (\ref{eq:single_copy_almost}) reduces to
		\begin{eqn}
			\partial_\nu F^{\mu \nu}
				&= g J^\mu,
		\end{eqn}
	which we recognize as Maxwell's equations.

To see the significance of the sign choice consider a Kerr-Schild spacetime sourced by a stationary pressureless fluid in four dimensions with a small energy density $\rho$, in which case the stress-energy tensor is $T_{\mu \nu} = \rho U_\mu U_\nu$ with $U^\mu \partial_\mu = (-g_{00})^{-1/2} \partial_0$. Then with mixed indices we have ${T^0}_0 = - \rho$, ${T^i}_0 = {T^i}_j = 0$, and hence
		\begin{eqn}
			J^0
				&= \pm \rho,
		&
			J^i 
				&= 0.
		\end{eqn}
So if we choose $k^0 = +1$ then the double copy maps positive charge
densities to positive masses and if we choose $k^0 = -1$ then the
double copy maps negative charge densities to positive
masses~\footnote{Strictly speaking, the single copy maps a gravity
  theory to a non-Abelian gauge theory, in which case the nature of
  the charge is more complicated. However, colour charge is removed
  upon taking the double copy, which explains why a positive mass
  always results.}. We will choose $k^0 = +1$ so that we can identify
the Kerr-Schild scalar field $\phi$ as the scalar potential on the
gauge side.
	
However, keep in mind that on the gravity side the choice of $k^0$ is entirely artificial - given a Kerr-Schild graviton $k_\mu k_\nu \phi$ we can always choose $k^0 = \pm 1$ by rescaling $k^\mu \mapsto \pm k^\mu / k^0$ and $\phi \mapsto (k^0)^2 \phi$. These redefinitions leave our spacetime unchanged but alter the gauge field with which it is identified. In flat space, choosing $k^0 = \pm 1$ is enough to get a reasonable single copy; in $d>4$ this means mapping point sources to point sources. When considering curved space, this choice can be more complicated (see \cite{Carrillo-Gonzalez:2017iyj} for a detailed analysis).

\section{2+1 Dimensional Gravity}
\label{sec:2+1}

General relativity in four spacetime dimensions is governed by the familiar Einstein-Hilbert action,
		\begin{eqn}
			S_\text{EH}
				= \int \text{d}^4 x \, \sqrt{-g} \bigg\{ \frac{1}{\kappa^2} (R - 2 \Lambda) 
					+ \mathcal{L}_\text{matter} \bigg\},
		\end{eqn}
	with $g = \det g_{\mu \nu}$, $R$ the scalar curvature, $\Lambda$ the cosmological constant, and $\mathcal{L}_\text{matter}$ the Lagrangian for the matter fields present. Variation with respect to the metric gives the Einstein field equations
		\begin{eqn}\label{eq:einstein_field_equations}
			G_{\mu \nu} + g_{\mu \nu} \Lambda
				&= \frac{\kappa^2}{2} T_{\mu \nu},
		\end{eqn}
where 
\begin{displaymath}
G_{\mu \nu} = R_{\mu \nu} - \frac{1}{2} g_{\mu \nu} R
\end{displaymath}
is the
Einstein tensor (with $R_{\mu \nu}$ the Ricci tensor) and 
\begin{displaymath}
T_{\mu \nu}
= -\frac{2}{\sqrt{-g}}\frac{ \delta (\sqrt{-g} \mathcal{L}_\text{matter})} 
{\delta g^{\mu \nu}} 
\end{displaymath}
the stress-energy tensor. We can
straightforwardly generalize this to an arbitrary number of spacetime
dimensions $d$ by
		\begin{eqn}
			S_\text{EH}
				= \int \text{d}^d x \, \sqrt{-g} \bigg\{ \frac{1}{\kappa^2} (R - 2 \Lambda) 
					+ \mathcal{L}_\text{matter} \bigg\},
		\end{eqn}
which leaves Eq. (\ref{eq:einstein_field_equations})
unchanged in form.

\subsection{Vacuum solutions in three spacetime dimensions are flat}

To see why setting $d = 3$ leads to a {\em
  qualitatively} different theory consider the following argument,
courtesy of \cite{Giddings:1983es}. In $d$ spacetime dimensions the
Riemann tensor ${R^\lambda}_{\mu \sigma \nu}$ has
\begin{displaymath}
N_\text{R} = \frac{1}{12} d^2 (d^2
- 1) 
\end{displaymath}
independent components, while the Einstein tensor $G_{\mu \nu}$
has 
\begin{displaymath}
N_\text{E} = \frac{1}{2} d (d+1). 
\end{displaymath}
It follows that for $d \geq 4$ we have $N_\text{R} > N_\text{E}$
(i.e. $d = 4$ gives $N_\text{R} = 20$ and $N_\text{E} = 10$). Every
component of $G_{\mu \nu}(x)$ is specified at every spacetime point
$x$ by the local matter distribution $T_{\mu \nu}(x)$ via the Einstein
field equations (\ref{eq:einstein_field_equations}), but with
$N_\text{R} > N_\text{E}$ this cannot uniquely fix every component of
${R^\lambda}_{\mu \sigma \nu}(x)$: given $G_{\mu \nu}(x)$ the system
of equations
		\begin{eqn}\label{eq:einstein_and_riemann}
			G_{\mu \nu} 
				= g^{\lambda \sigma} R_{\lambda \mu \sigma \nu} 
					- \frac{1}{2} g_{\mu \nu} g^{\lambda \sigma} g^{\alpha \beta} 
						R_{\lambda \alpha \sigma \beta}
		\end{eqn}
which relates ${R^\lambda}_{\mu \sigma \nu}(x)$ to $G_{\mu \nu}(x)$ is
a set of $N_\text{E}$ equations in $N_\text{R}$ unknowns and is hence
underdetermined. It is precisely this freedom which allows for all the
familiar gravitational phenomena we know and love: waves can propagate
through the vacuum, two objects separated by empty space can gravitate
toward each other, etc.

Let us now consider $d = 3$ spacetime dimensions, where from the above
results we have $N_\text{R} = N_\text{E} = 10$. This
implies that Eq. (\ref{eq:einstein_and_riemann}) can be inverted, and
invoking (\ref{eq:einstein_field_equations}) we find
		\begin{eqn}
			R_{\lambda \mu \sigma \nu}
				&= \frac{\kappa^2}{2} \bigg\{ g_{\lambda \sigma} T_{\mu \nu}
						+ g_{\mu \nu} T_{\lambda \sigma}
						- g_{\lambda \nu} T_{\mu \sigma}
						- g_{\mu \sigma} T_{\lambda \nu}
						+ T (g_{\lambda \nu} g_{\mu \sigma}
							- g_{\lambda \sigma} g_{\mu \nu}) \bigg\}.
		\end{eqn}
In other words, the curvature ${R^\lambda}_{\mu \sigma \nu}$ at each
point is determined entirely by $T_{\mu \nu}$ {\em at that same
  point}. In particular, the condition $T_{\mu \nu}(x) = 0$ at a given
spacetime point implies
		\begin{eqn}
			{R^\lambda}_{\mu \sigma \nu}(x)
				= 0 \ ,
		\end{eqn}
such that no matter how much energy is contained in a given region of
space, that energy will have no impact on the curvature of spacetime
anywhere else. More generally, given a region $S$, the geodesics
through any other region are completely insensitive to the energy 
and momentum distribution in $S$.

An alternate way to view this is that the graviton is
overconstrained. As is well-known, the physical degrees of freedom of
the graviton $h_{\mu \nu}$ can be obtained from the transverse and
traceless part of the spatial components $h_{ij}$ of the metric. In $d$
spacetime dimensions these both amount to $d-1$ separate constraints,
which if $d \geq 4$ cannot completely determine the
$\frac{1}{2}d(d-1)$ independent components of a symmetric
$(d-1)$-dimensional rank-two tensor. If $d = 3$ on the other hand, we
have $2(d-1) = 4$ separate constraints on the $\frac{1}{2}d (d-1) = 3$
independent components of $h_{ij}$, and are led to the conclusion that
the graviton does not propagate in $d = 3$.

\subsection{Three-dimensional gravity has no Newtonian limit}
\label{sec:no_newtonian_limit}

We will here take the Newtonian limit of pure gravity in arbitrary dimensions
and show that in $d=3$ the Newtonian potential decouples from any
matter sources present.

We take the Newtonian limit as follows. We consider weak fields, i.e. fields for which
there exists a coordinate system in which the metric is 
quasi-stationary, $\partial_0 g_{\mu \nu} \approx 0$, and only weakly 
deviates from flat space, $g_{\mu \nu} = \eta_{\mu \nu}
+ h_{\mu \nu}$ with $|h_{\mu \nu}| \ll 1$. (In this section, we 
reabsorb the factor of $\kappa$ into $h_{\mu \nu}$.) We consider nonrelativistic
trajectories, i.e. geodesics $x^\mu(\tau)$ for which $|\dot{x}^i(\tau)| \ll 1$ in this
coordinate system, as well as a stationary weak source, i.e. a
stress-energy tensor with ${T^0}_0 = -\rho$ small and every other component 
vanishing. Given this, we neglect all terms beyond linear order in $\rho$ and
$h_{\mu \nu}$.

It may then be shown that the spatial geodesic equation reduces 
to
		\begin{eqn}\label{eq:geodesic_newtonian}
			\frac{\text{d}^2 \bs{x}}{\text{d}t^2} 
				= \frac{1}{2} \vec{\nabla} h_{00}
				\equiv - \vec{\nabla} \Phi,
		\end{eqn}
where we identify
		\begin{eqn}
			\Phi
				\equiv - \frac{1}{2} h_{00}
				= - \frac{1}{2}(1 + g_{00})
		\end{eqn}
as the Newtonian potential, and that the $00$-component of the
trace-reversed Einstein equations becomes in $d$ dimensions
		\begin{eqn}\label{eq:poisson_arbitrary_d}
			\vec{\nabla}^2 \Phi
				&= \frac{\kappa^2}{2} \rho \bigg( \frac{d-3}{d-2} \bigg).
		\end{eqn}
It is noteworthy to mention that the explicit expressions for both the Newton and Poisson equations in the  Newtonian limit are gauge dependent. For example, had we worked on De Donder gauge, we would have obtained that the  dimension-dependent factor appears in Newton's equation instead of in Poisson's equation \cite{garcia-diaz_2017}.

Specializing to a point mass $M$ at the origin we then find the Newtonian
gravitational field $\mathbf{g} = \vec{\nabla} \Phi$ a distance $r$ from the 
origin to be given by
		\begin{eqn}
			\mathbf{g}(r)
				= - \mathbf{\hat{r}} \bigg( \frac{M}{r^{d-2}} \bigg)
					\frac{\kappa^2}{2}
					\left\{ \frac{(d-3)\Gamma \big(\frac{d+1}{2} \big)}
						{(d-1)(d-2) \pi^{(d-1)/2}} \right\}.
		\end{eqn}
From this last equation we can then read off the $d$-dimensional Newton constant
$G_d$:
		\begin{eqn}
			G_d
				&= \frac{\kappa^2}{2}
						\left\{ \frac{(d-3)\Gamma \big(\frac{d+1}{2} \big)}
							{(d-1)(d-2) \pi^{(d-1)/2}} \right\}.
		\end{eqn}
In $d=4$ this yields (via $\Gamma(5/2) = 3 \sqrt{\pi}/4$) the expected result
$G_4 = \frac{\kappa^2}{16 \pi}$. However in $d=3$ we obtain $G_3 = 0$. 
Hence, in three spacetime dimensions the Newtonian gravitational field
 $\mathbf{g}$ in the presence of a point mass (or any other matter distribution, as seen from Eq.\eqref{eq:poisson_arbitrary_d}) vanishes identically
at any distance.  The Newtonian potential $\Phi$ is no longer coupled via 
Poisson's equation (\ref{eq:poisson_arbitrary_d}) to the energy distribution $\rho$,
and nonrelativistic trajectories (\ref{eq:geodesic_newtonian}) are straight lines.

In this section, we have reviewed salient properties of three
dimensional gravity, in particular the absence of propagating graviton
degrees of freedom. As discussed in the introduction, this creates a
potential problem for the classical double copy for exact solutions,
given that the latter maps gauge boson degrees of freedom to the
graviton in all known
cases~\cite{Monteiro:2014cda,Luna:2015paa,Luna:2016due,Berman:2018hwd}. To
investigate this further, we first review a known three-dimensional
gravitational system, and its single
copy~\cite{Carrillo-Gonzalez:2017iyj}.

\section{The BTZ Black Hole}
\label{sec:BTZ}

Since gravity does not propagate in three dimensions it was long
thought that there were no related black hole solutions of Einstein's
equations. In 1992 Ba{\~n}ados, Teitelboim, and Zanelli showed
\cite{Banados:1992wn} that in fact there do exist black hole
solutions, collectively referred to as the {\it BTZ black hole}, in
asymptotically AdS spacetimes. Geometrically, this black hole solution
can be constructed as a quotient of the covering of AdS$_3$ by a
discrete group of isometries \cite{Banados:1992gq}.

In the following we review the BTZ black hole in the context of the Kerr-Schild double copy which was first analyzed in \cite{Carrillo-Gonzalez:2017iyj}. For clarity, we will follow the conventions on that paper which correspond to
\begin{align}
g_{\mu\nu}=\eta_{\mu\nu}+k_\mu k_\nu\phi \ , \\
\kappa^2\rightarrow g  \ , \quad  M\rightarrow Q \ .
\end{align}
In the rest of the paper we stick to the conventions stated previously.

\subsection{Features}

Importantly for our purposes, the BTZ spacetime admits a Kerr-Schild
description~\cite{Kim:1998iw}. More specifically, one may adopt a
spheroidal coordinate system, such that the BTZ solution has the form
of Eq.~(\ref{KSdef}), with the following results for the Minkowski
line element, Kerr-Schild vector and scalar function respectively:
		\begin{eqn}\label{eq:BTZ_kerr-schild}
			\eta_{\mu \nu} \text{d}x^\mu \text{d}x^\nu
				&= - \text{d}t^2 + \frac{r^2}{r^2 + a^2} \text{d}r^2
					+ (r^2 + a^2) \text{d}\theta^2,
		\\
			k_\mu \text{d}x^\mu
				&= - \text{d}t + \frac{r^2}{r^2 + a^2} \text{d}r - a \, \text{d}\theta,
		\\
			\phi
				&= 1 + \frac{\kappa^2 M}{\pi} - \frac{r^2}{\ell^2}.
		\end{eqn}
Here $\Lambda = - 1/\ell^2$ is the cosmological constant, $M$ a
mass-like parameter (the charge associated with the time translation
symmetry of the spacetime), and $a$ an angular momentum-like parameter
(the charge associated with the rotational symmetry of the
spacetime). This is reminiscent of the Kerr solution in
$d=4$. However, unlike in the latter case, the spacetime is {\em not}
sourced by a rotating disc of finite radius: it cannot be, due to the
fact that gravity does not propagate in three spacetime dimensions. In
fact, the above metric describes different spacetimes depending on the
values of $M$ and $a$. For $M>0$ and $|a|<\ell$ we have a black hole
solution that shares many features with more familiar four-dimensional
black holes, the most obvious being a horizon, which turns out to be
located at
		\begin{eqn}
			r_\text{horizon}
				= \ell \sqrt{\frac{M}{2} \bigg(1 + \sqrt{1 - (a / \ell)^2} \bigg)}.
		\end{eqn}
We note also that in the limit $a,M \to 0$ we obtain (with a
coordinate change)
		\begin{eqn}
			\text{d}s^2
				= - \bigg( \frac{r}{\ell} \bigg)^2 \text{d}t^2
					+ \bigg( \frac{r}{\ell} \bigg)^2 \text{d}r^2
					+ r^2 \text{d} \theta^2,
		\end{eqn}
	which is {\em not} the $\text{AdS}_3$ vacuum we would expect to find in the massless limit. Rather we obtain $\text{AdS}_3$ with $a = 0$ and $M = -1$,
		\begin{eqn}
			\text{d}s^2
				= - \bigg( 1 + \bigg( \frac{r}{\ell} \bigg)^2 \bigg) \text{d}t^2
					+ \bigg( 1 + \bigg( \frac{r}{\ell} \bigg)^2 \bigg)^{-1} \text{d}r^2
					+ r^2 \text{d} \theta^2,
		\end{eqn}
	although every other negative value of $M$ is unphysical due to a naked conical singularity \cite{Banados:1992wn}.

\subsection{The BTZ single copy}

Since we can write the BTZ black hole in Kerr-Schild form, we can
obtain its single copy, as first presented in
ref.~\cite{Carrillo-Gonzalez:2017iyj}. That is, we may construct a
gauge field $A_\mu$ such that
		\begin{eqn}
			A_\mu \text{d}x^\mu
				= \phi k_\mu \text{d}x^\mu
				= \bigg( 1 + \frac{g \ Q}{\pi} + \rho \, r^2\bigg)
					\bigg( - \text{d}t + \frac{r^2}{r^2 + a^2} \text{d}r 
						- a \, \text{d}\theta \bigg),
		\end{eqn}
where we have relabelled $\Lambda\rightarrow \rho$.  Substituting this into the Maxwell equations of Eq.~(\ref{Maxwell}), one finds that the source is a constant charge density filling all space: 
		\begin{eqn}
			J^\mu
				&= 4 \rho v^\mu,
		&
			v^\mu \partial_\mu
				= \partial_t,
		\end{eqn}
 This is consistent
with the single copy of four-dimensional (Anti-)de Sitter space in
ref.~\cite{Luna:2015paa}, which replaces the constant energy density
in the gravity theory with a constant charge density in the gauge
theory. The mass parameter $M$ for the BTZ black hole disappears in
the gauge theory source: it corresponds to a constant contribution to
the electrostatic potential on the gauge theory side, which has no
dynamical consequences. This is in stark contrast to the Schwarzschild
black hole in four dimensions, in which the point mass $M$ on the
gravity side is associated with a point charge in the gauge
theory. This should not surprise us, given that there is no known
analogue of the Schwarzschild black hole in $d=3$. On the contrary,
the unusual role of the mass parameter in the single copy serves as a
further illustration of how different the BTZ black hole is to the
usual Schwarzschild or Kerr black holes living in four-dimensional
Minkowski space. One can also notice that; while  the BTZ  metric can describe  a  black hole, a particle giving rise to a conical singularity, or simply AdS$_3$; the gauge solution does not correspond to physically different solutions when varying the free parameters $Q$ and $\rho$. 
	
In this section, we have reviewed the well-known BTZ black hole in
$d=3$ spacetime dimensions~\cite{Banados:1992gq}, which has a
well-defined gauge theory single
copy~\cite{Carrillo-Gonzalez:2017iyj}, whose interpretation (in terms
of the cosmological constant) is consistent with previous results in
four spacetime dimensions. We should perhaps not be at all surprised
that the single copy works in this case, given that the potential
difficulties of three spacetime dimensions are on the gravity side of
the double copy correspondence, rather than the gauge theory side. If
we start with a sensible gravity solution, we then expect a sensible
gauge theory counterpart. A trickier situation would be to start with
a well-defined gauge theory object, and to seek its gravitational
analogue. The simplest such object is a pointlike electric charge,
which we will now investigate further.

\section{The double copy of a point charge}
\label{sec:pointcharge}

In all of the above examples, we have started with a gravity solution
and obtained its gauge theory counterpart via the single copy. In this
section, we will reverse this procedure, by starting from a gauge
field and finding its corresponding Kerr-Schild double copy. This
allows us to directly probe the issue of how the physical degrees of
freedom in the gauge theory get mapped under the double copy. Since we
will be working in (2+1) dimensions, there will be no graviton degrees
of freedom; instead, as already discussed in section~\ref{sec:intro},
we expect to turn on the dilaton degree of freedom. Furthermore, the
lack of a graviton is related to the well-known absence of a
well-behaved Newtonian limit for three-dimensional General
Relativity~\cite{Giddings:1983es}, although there are no such issues
in the gauge theory. Let us examine this issue in more detail.

\subsection{Expecting Newtonian Behavior}
\label{sec:expecting_newton}

Suppose we have some stationary gauge field $A_\mu$. In order to take
its Kerr-Schild double copy we need to specify a splitting
\begin{displaymath}
A_\mu =\phi k_\mu,
\end{displaymath}
for some scalar function $\phi$, where $k_\mu$ is null and
geodesic\footnote{Note that we can only take the double copy in
certain gauges, since if we choose e.g. Coulomb gauge, $A_0 = 0$,
then there exists no null vector $k_\mu$ such that $A_\mu = \phi
k_\mu$. This mimics a similar behaviour in amplitudes, namely that
the double copy correspondence is manifest only in certain
(generalised) gauges~\cite{Bern:2010ue}. For a recent discussion
beyond this, see ref.~\cite{Bern:2017yxu}.}. Choosing $k^0 = 1$ 
we identify $A^0 = \phi$, and the timelike Maxwell equation becomes
		\begin{eqn}\label{eq:scalar_poisson}
			g J^0
				= \partial_\mu F^{0 \mu}
				= \partial_\mu (\partial^0 A^\mu - \partial^\mu A^0)
				= - \partial^2 A^0
				= - \partial^2 \phi
				= - \vec{\nabla}^2 \phi,
		\end{eqn}
using the stationarity of $A_\mu$ throughout. Thus, the scalar $\phi$
satisfies Poisson's equation with source $-J^0$. The field $\phi$ is
thus also described by the linearised equations of {\it biadjoint
scalar theory}, and can then be interpreted as the so-called {\it
zeroth copy} of the gauge field $A_\mu$~\cite{Monteiro:2014cda}.

Let us now consider the Newtonian limit of the double copy of this 
gauge field. With $k^0 = 1$ the purely timelike component of the metric is
		\begin{eqn}
			g_{00}
				= -1 + \kappa \phi,
		\end{eqn}
so the Kerr-Schild field $\phi$ is related to the Newtonian potential 
$\Phi$ as
		\begin{eqn}\label{eq:relate_the_phis}
			\Phi 
				= - \frac{1}{2} (1 + g_{00}) 
				= - \frac{\kappa}{2} \phi.
		\end{eqn}
Furthermore, we can relate $\Phi$ to the source on the gravity side using
eqs. (\ref{eq:single_copy_source}) and (\ref{eq:scalar_poisson}) and making 
the usual substitution $\frac{\kappa}{2} \leftrightarrow g$:\footnote{This is
the more general form of the relation (\ref{eq:poisson_arbitrary_d}), which we
obtain from Eq.(\ref{eq:double_copy_poisson}) 
by taking the Newtonian limit of the source. (The second line of Eq. (\ref{eq:double_copy_poisson}) 
can be derived without reference to the double copy, but
doing it this way clarifies the relationship between the Newtonian and 
electrostatic potentials and their sources.)}
		\begin{align}\label{eq:double_copy_poisson}
			\vec{\nabla}^2 \Phi
				&= \frac{\kappa^2}{4} J^0\notag\\
				&= \frac{\kappa^2}{2} \bigg( \frac{T}{d-2} - {T^0}_0 \bigg).
		\end{align}
This is Poisson's equation for the Newtonian potential $\Phi$, and
tells us that any metric obtained by double-copying a gauge solution
should have a Newtonian potential sourced by the timelike component
of the gauge source, modulo replacements of charges by masses and the
gauge coupling constant with its gravitational counterpart.\footnote{Note
 that the minus sign in Eq. (\ref{eq:relate_the_phis})
is necessary for our results to be consistent. We have already claimed
 that the choice $k^0 = +1$ allows
us to identify positive charges with positive masses. The negative sign
in $\Phi \propto - \phi$ is crucial for this identification to hold: with positive
 $\rho = J^0$ the solutions of $\vec{\nabla}^2 \phi \propto - J^0$ are
 repulsive while those of $\vec{\nabla}^2 \Phi \propto + J^0$ are 
 attractive.}

The above arguments make sense in $d \geq 4$, given the existence of a
well-defined Newtonian limit for gravity. Even in $d = 3$ we can
see that in the case of the BTZ black hole our results
 accord exactly with what we've already found: applying Gauss'
law to a constant energy distribution (i.e. a cosmological constant)
in two dimensions yields a Newtonian gravitational field satisfying
$|\mathbf{g}| \propto r$, which corresponds to a potential satisfying
$|\Phi| \propto r^2$, in agreement with
Eq. (\ref{eq:BTZ_kerr-schild}). 

In general in the $d = 3$ case Eq. 
(\ref{eq:double_copy_poisson}) becomes
 		\begin{eqn}\label{eq:dc_poisson_3d}
			\vec{\nabla}^2 \Phi
				= \frac{\kappa^2}{4} J^0
				= \frac{\kappa^2}{2}(T - {T^0}_0).
		\end{eqn}
That three-dimensional gravity has no Newtonian limit is down
to the fact that if we take the Newtonian limit of the source
as in sec. (\ref{sec:no_newtonian_limit}) then $T = {T^0}_0
= - \rho$ and the right hand side identically vanishes.
From the perspective of the double copy this implies that
the identification of a small charge density $J^0$ on the gauge side
with a small energy density $T_{00}$ on the gravity side, which
holds in $d \geq 4$, breaks down in three dimensions. Hence it 
becomes natural to ask: what gravitational source {\em does} 
correspond to a small charge density in three dimensions?

As we can see from Eq. (\ref{eq:dc_poisson_3d}), such a source
must have nonzero spatial diagonal components even in this quasi-Newtonian limit  in order for $T - {T^0}_0$ to be non-vanishing.
Further, since we know that a nontrivial scalar potential $\phi$ will 
generically be non-vanishing everywhere,  we expect that the double copy will have nonzero curvature. Since three 
dimensional gravity does not propagate, this implies that
{\em the double copy of a localized charge density in three dimensions
cannot be a vacuum solution}, but rather the gravitational source 
must permeate spacetime in such a way as to mimic the Newtonian
behavior of higher-dimensional gravity. In what follows we 
present an explicit example of such a system.

%We thus conclude that the double copy of any three-dimensional
%gauge theory solution
%  with a localized source cannot be a vacuum solution of Einstein's
%  equations. Furthermore, the presence of a non-zero source on the
%gravity side can mimic the presence of a Newtonian potential, that
%would otherwise be zero in pure gravity. 

\subsection{Point Charge in 2+1 Dimensions and   its Double  Copy}
\label{sec:chargeDC}

As a specific example of the Kerr-Schild double copy of a given gauge
field solution, we now consider a static point charge sitting at the
origin. The corresponding gauge field can be written as
		\begin{eqn}
			A_\mu
				&= k_\mu \phi,
		&
			k_\mu \text{d}x^\mu
				&= - \text{d}t - \text{d}r,
		&
			\phi
				&= - \frac{ g Q }{2\pi} \log{r},
		\end{eqn}
using polar coordinates $(t,r,\theta)$.
One may explicitly verify that the gauge field $A^\mu$ 
satisfies the Maxwell equations
$\nabla_\nu F^{\mu \nu}=\,J^\mu$ with a source current
		\begin{displaymath}
			J^\mu \partial_\mu
				=g Q \delta^2(r) \partial_t,
		\end{displaymath}
using the fact that $\vec{\nabla}^2 \log r = 2 \pi \delta^2(r)$
in two spatial dimensions. Strictly speaking, we should introduce 
a scaling factor $r_0$ into $\phi$ so as to make the logarithm 
dimensionless. However, we may choose units such that 
this is unity for what follows.  Notice that
the vector $k$ is null and geodesic as required,\footnote{Our 
gauge field, which has nonzero spatial components, is related 
to the perhaps more intuitive solution 
$A^0 = - \frac{g Q}{2 \pi} \log r$, $A^i = 0$ by
a gauge transformation, which is necessary in order for $k$ to
satisfy these requirements. (See the previous footnote
about the gauge-dependence of the double copy.)} and $\phi$ satisfies
the linearized (static) equations of motion of the biadjoint
scalar. Given this, we are able to construct the classical double copy
of the static charge. The double copy is given by the Kerr-Schild
metric
\begin{equation}
g_{\mu\nu}=\eta_{\mu\nu}+ \kappa \,  k_\mu  k_\nu \phi=\left(
\begin{array}{ccc}
-1- \frac{\kappa^2 M}{4 \pi}  \log{r} & - \frac{\kappa^2 M}{4 \pi} \log{r}  &0\\
- \frac{\kappa^2 M}{4 \pi}   \log{r}  & 1- \frac{\kappa^2 M}{4 \pi} \log{r} & 0 \\
0 & 0 & r^2 \\
\end{array}
\right) \ , \label{metric}
\end{equation}
where we have made the standard
replacements~\cite{Monteiro:2014cda}
\begin{equation}
g\rightarrow\frac{\kappa}{2} \ , \quad Q \rightarrow M \ .
\end{equation}
Substituting the metric of Eq.~(\ref{metric})
into the Einstein's equations, we find a non-zero stress tensor given
by
\begin{eqnarray}
{T^\mu}_\nu=-\frac{1}{\kappa} \left(
\begin{array}{ccc}
\frac{\phi'}{r}& 0 & 0 \\
0 & \frac{\phi'}{r}& 0 \\
0 & 0 & \phi'' \\ 
\end{array} 
\right)=\frac{M}{4\pi\,r^2} \left(
\begin{array}{ccc}
1& 0 & 0 \\
0 &1& 0 \\
0 & 0 & -1\\
\end{array}
\right)  \ . \label{stresstensor}
\end{eqnarray}
It is interesting already to note that, although the gauge field
source is localized to the origin, the gravitational source is
not. This is possible due to the relation between the sources given in
Eq.\eqref{eq:single_copy_source}, and indeed one can easily see that
this relation holds: using the first expression for the stress tensor
in Eq. \eqref{stresstensor} leads to a gauge theory source current
\begin{displaymath}
J^\mu \partial_\mu
	= (- \vec{\nabla}^2\phi,0,0)=gQ\,\delta^2(r)(1,0,0) , 
\end{displaymath}
as required. We have therefore found an explicit example of the
conclusion of the previous section, namely that the double copy of a
localised gauge theory source cannot correspond to a vacuum solution
of the Einstein equations in $d=3$. 

The presence of a non-trivial source term implies the presence of
matter degrees of freedom. This may at first sight seem puzzling,
until we remember that the double copy of pure gauge theory is not
pure gravity, but gravity coupled to a dilaton and two-form. The
latter will be absent, as in previous
examples~\cite{Monteiro:2014cda,Luna:2015paa,Luna:2016due,Berman:2018hwd},
due to the symmetric nature of the double copied field. However, the
dilaton may indeed be turned on, and thus we would like to identify
the above stress-energy tensor with a non-zero dilaton field. In order
to do so, we consider a free massless scalar field. Such a field is
invariant under the symmetry 
\begin{equation}
\varphi\rightarrow\pm\varphi+c,
\label{phisym} 
\end{equation}
and its stress-energy tensor reads
\begin{equation}
{T^\mu}_\nu=\partial^\mu\varphi\partial_\nu\varphi-\frac{1}{2}g^\mu_\nu(\partial\varphi)^2 \ .
\end{equation} 
The Einstein equations then imply a Ricci tensor 
\begin{equation}
{R^\mu}_{\nu}=\kappa^2 /2 \, \partial^\mu\varphi\partial_\nu\varphi.
\label{dilaton_ricci}
\end{equation}
On the other hand, from the metric in Eq.\eqref{metric} we find
\begin{equation}
{R^\mu}_{\nu}=-\frac{\kappa^2\,M}{4\pi r^2}\left(
\begin{array}{ccc}
0& 0 & 0\\
0  & 0 & 0 \\
0 & 0 &1 \\
\end{array}
\right) \ . \label{ricci}
\end{equation}
Comparing these results, we obtain that the dilaton profile must be
linear in the azimuthal angle. Furthermore, the difference in sign
between eqs.~(\ref{dilaton_ricci}) and~(\ref{ricci}) implies that
either the scalar field is imaginary, or that it has the wrong sign in
its kinetic term. The second option in fact makes sense, albeit for a
rather complicated reason. For amplitudes, the double copy is known to
be true only in certain {\it generalised gauges}, where one has
potentially performed both a gauge transformation and a field
redefinition. It was known already before the double
copy~\cite{Bern:1999ji} that it is possible to make field
redefinitions so as to make the gravitational Lagrangian consistent
with the KLT relations~\cite{Kawai:1985xq} relating gauge and gravity
amplitudes at tree-level (and which are equivalent to the double
copy). That is, by introducing a dilaton and starting in the de Donder
gauge, one may transform the graviton and dilaton fields so as to
remove all the terms in the graviton Lagrangian that mix left and
right indices, leaving a structure that is manifestly
double-copy-like. This turns out to have the side-effect of reversing
the sign of the dilaton kinetic term. Returning to our present study
of classical solutions, we should then expect that if the double copy
of a given gauge field generates a non-zero dilaton profile, the
latter should be associated with a ``wrong-sign'' dilaton kinetic
term. The fact that this has not been noticed before is simply because
previous examples of the Kerr-Schild double copy involve zero dilaton
profiles on the gravity
side~\cite{Monteiro:2014cda,Luna:2015paa,Luna:2016due,Berman:2018hwd}. From
the above results, the explicit form of the dilaton for the double
copy of the point charge in $d=3$ is defined by\footnote{The reader
  may worry that the form of the dilaton itself is linear in the
  azimuthal angle, and thus not periodic. However, the Einstein
  equations imply that Eq.~(\ref{derphi}) need only be satisfied {\it
    locally}. A global dilaton profile can be obtained by patching
  solutions together, up to transformations of the form of
  Eq.~(\ref{phisym}).}
\begin{equation}
\partial_\mu \varphi=\sqrt{M/2\pi} \, (0,0,1) \ . \label{derphi}
\end{equation}
The identification of the source term in Eq.~(\ref{stresstensor}) with
a nonzero dilaton profile is natural, given that a similar matching of
gauge theory degrees of freedom with the dilaton can be made for the
plane wave states that enter the original BCJ double copy for
scattering amplitudes. However, we may still worry whether or not this
is the only possible interpretation of Eq.~(\ref{stresstensor}). To
this end, two remarks are useful. Firstly, we may check the equation of
motion of the dilaton. Considering the field profile from
Eq.\eqref{derphi} we see that it indeed satisfies
\begin{equation}
\nabla^\mu\nabla_\mu\varphi=0 \ ,
\end{equation}
where $\nabla_\mu$ is the covariant derivative of the metric
\eqref{metric}. Secondly, we have verified that the stress tensor from
Eq.\eqref{stresstensor} cannot correspond to that of a standard
viscous fluid, which reinforces its dilaton interpretation; for details
on this calculation see Appendix~\ref{nofluid}. In the following
subsections, we analyze different aspects of this dilaton-gravity
solution which will strengthen its interpretation as the double copy
of a static charge in (2+1) dimensions.

\subsubsection{Generalized gauge and weak energy conditions}

Above, we have seen that interpreting the gravity source of
Eq.~(\ref{stresstensor}) in terms of a dilaton implies a ``wrong
sign'' kinetic term for the latter. For the coupled dilaton / gravity
system to be consistent, it must be possible to perform a generalised
gauge transformation such that both the dilaton and graviton have the
correct signs in their respective kinetic terms. In such a generalised
gauge, the double copy will no longer be manifest, and the field
equations will no longer be linearised in general (i.e. the
generalised gauge transformation may be highly non-linear). In order
to inspect the kinetic terms, however, we may simply restrict
ourselves to generalised gauge transformations at linear order. We may
then proceed to a ``correct sign'' generalised gauge in two steps. We
may first gauge transform the graviton of Eq.~(\ref{metric}) to the de
Donder gauge defined by
\begin{displaymath}
\partial_\mu\bar{h}^{\mu\nu}=0,\quad 
\bar{h}_{\mu\nu}=h_{\mu\nu}-\frac{1}{2}\eta_{\mu\nu}h^{\alpha\beta}h_{\alpha\beta}.
\end{displaymath}
This gauge transformation corresponds to the linear diffeomorphism
\begin{displaymath}
h_{\mu\nu}\rightarrow h_{\mu\nu}+\partial_\mu \xi_\nu+\partial_\nu
\xi_\mu,
\end{displaymath}
where we find the following explicit forms for the components of
$\xi_\mu$:
\begin{equation}
\xi_t= \frac{\kappa^2}{4 \pi} r( \log{r}-1) \ ,  \quad \xi_r= \frac{\kappa^2}{8 \pi} r( \log{r}-1) \  ,  \quad  \xi_\theta=0    .
\end{equation}
In this gauge, the quadratic part of the Lagrangian reads
\begin{equation}
\lag_2=-\frac{1}{2}\partial_\alpha h_{\mu\nu}\partial^\alpha h^{\mu\nu}+\frac{1}{4}\partial_\alpha h_{\mu}^\mu\partial^\alpha h^{\nu}_\nu+\frac{1}{2}\partial_\alpha\varphi\partial^\alpha\varphi \ .
\end{equation}
Next, we may perform the field redefinitions
\begin{equation}
h_{\mu\nu}\rightarrow h_{\mu\nu}-\eta_{\mu\nu} A \, \varphi \ , \qquad \varphi\rightarrow-\frac{1}{\sqrt{2}}h_{\mu}^{\mu}+\sqrt{\frac{3A^2}{2}-1} \, \varphi  \ ,  \label{redef}
\end{equation}
where $A$ is a constant subject to the constraint
$A>\sqrt{\frac{2}{3}}$, to obtain
\begin{equation}
\lag_2=-\frac{1}{2}\partial_\alpha h_{\mu\nu}\partial^\alpha h^{\mu\nu}-\frac{1}{2}\partial_\alpha\varphi\partial^\alpha\varphi \ .
\end{equation}
In this last expression, both the dilaton and graviton have the
correct sign for the kinetic term, thus showing that a ``right sign''
Lagrangian can indeed be obtained from our coupled dilaton-gravity
system by a generalised gauge transformation.

If we simply analyze the gravitational theory whose stress-tensor is
given by Eq.\eqref{stresstensor} (i.e. with the wrong-sign dilaton),
we will reach the conclusion that the theory contains a ghost field
that renders the vacuum unstable to processes such as $0\rightarrow
hh+\varphi\varphi$. In other words, the stress tensor in
Eq.\eqref{stresstensor} does not satisfy the Weak Energy Condition,
which would tell us that the energy of the dilaton is negative. In
this setting, it is possible to think of the theory as an effective
field theory with a momentum cutoff; in such case, higher dimensional
operators could render the theory unitary
\cite{Carroll:2003st,Cline:2003gs}. In the present case, we will
interpret the situation in a different way. When obtaining the double
copy, the gravitational theory is in a generalized gauge where matter
and geometry are mixed. Once we switch to de Donder gauge and perform
the field redefinitions in Eqs.\eqref{redef}, we obtain a theory where
both the graviton and dilaton have the correct kinetic term signs. In
this generalized gauge, the weak energy condition is
satisfied\footnote{The issue of whether physical energy conditions in
  GR are satisfied in generalised gauges for which the double copy is
  satisfied has also been discussed previously (see
  e.g. Ref.~\cite{Ridgway:2015fdl}).}.

\subsubsection{Newtonian-like Limit}

In this section, we examine further aspects of the Newtonian-like limit of
our gravity-dilaton system. First, we recall that the trajectory
$x^\mu(t)$ of a particle interacting with Newtonian gravity in (2+1)
dimensions satisfies
		\begin{eqn}\label{eq:newtonian_trajectory}
			\frac{\ud^2 \bs{x}}{\ud t}
				&= - \vec{\nabla} \Phi,
		&
			\vec{\nabla}^2 \Phi 
				&= 2 \pi G \rho,
		\end{eqn}
where $\Phi$ is the Newtonian potential, $\rho$ is the mass
density, and $G$ is Newton's constant. Outside a spherical source 
the Newtonian potential is given
by $\Phi = G M \log{r}$.
On the other hand, the
Newtonian limit of General Relativity in $d$ dimensions gives rise
to
		\begin{eqn}\label{NewtLim}
			\frac{\ud^2 \bs{x}}{\ud t^2}
				&= - \vec{\nabla} \Phi,
		&
			\vec{\nabla}^2 \Phi
				&= \frac{\kappa^2}{2} \rho
					\bigg( \frac{d-3}{d-2} \bigg),
		\end{eqn}
as in eqs. (\ref{eq:geodesic_newtonian}, \ref{eq:poisson_arbitrary_d}), which as
previously mentioned shows that in (2+1) dimensions masses do not experience 
Newtonian forces.

There have been some attempts at reproducing the expected Newtonian
limit in scalar-tensor theories. For example, a solution with a
logarithmic Newtonian potential at short distances was found in
\cite{Verbin:1994ks}. There, a scalar-tensor theory in (2+1)
dimensions was formulated as a dimensional reduction of of a 4d
theory, by taking solutions with no dependence in a particular
direction. Similar approaches have also been examined in
e.g. ref.~\cite{Barrow_1986}. In this paper, we have seen that the
Kerr-Schild double copy naturally generates a scalar-tensor theory
that mimics the presence of Newtonian gravity:
Eq.~(\ref{eq:double_copy_poisson}) relates the relevant Newtonian
potential to the gravitational source generated by double-copying a
given gauge field, where the source is associated with a nonzero
dilaton. 

Let us now consider the Newtonian limit of the explicit
dilaton-gravity solution found above. This limit
is in fact not Newtonian in the strictest sense:
we assume nonrelativistic trajectories, so that we
obtain $\text{d}^2 \bs{x} / \text{d}t^2 = - \vec{\nabla} \Phi$ with
$\Phi = - \frac{1}{2} h_{00}$, but we do not take the usual
Newtonian limit of the source, since doing so
requires that the source term $\frac{\kappa^2}{2} (T - {T^0}_0)$
for $\Phi$ vanish. Indeed, if we just leave the gravitational source
alone and use Eq. (\ref{stresstensor}) we find
		\begin{eqn}\label{eq:poisson_dc}
			\vec{\nabla}^2 \Phi
				&= \frac{\kappa^2}{4} M \delta^2(r),
		\end{eqn}
from which we deduce the non-point-like form of the stress-energy tensor
(\ref{stresstensor}) is such that the corresponding Newtonian
potential has a pointlike source. Further, from the Newtonian potential
		\begin{eqn}
			\Phi
				&= \frac{\kappa^2 M}{8 \pi} \log r \label{NewtPot}
		\end{eqn}
	we can read off Newton's constant for our system:
		\begin{eqn}
			G
				&= \frac{\kappa^2}{8 \pi}.
		\end{eqn}

In obtaining Eq. (\ref{eq:poisson_dc}) we did not need to make any assumptions 
about the source, since from the mixed-index stress-energy 
tensor (\ref{stresstensor}) we can just take the trace and evaluate. However
from this equation it is clear that the appropriate limit to take is the one in which 
the parameter $GM$, which multiplies {\em all} diagonal elements
of the stress-energy tensor, is small.
 This corresponds to taking the charge $Q$ of the point source on the gauge side
to be small, but does {\em not} correspond to a
small point mass on the gravity side, since the gravitational
source is nonzero throughout spacetime. But by the arguments of sec. 
 \ref{sec:expecting_newton} and Eq. (\ref{eq:poisson_dc}) we see that this 
is in fact the correct limit to take if we wish to recover a Newtonian limit 
with a nonvanishing pointlike source, which requires a stress-energy
tensor which is nonvanishing throughout spacetime.

In simple terms, the lack of a 
propagating graviton upon taking the double copy forces the dilaton 
to turn on, in order to match up with the physical degrees of 
freedom in the gauge theory. This then ensures a well-behaved Newtonian-like limit, in
accordance with the fact that there is a well-behaved Newtonian limit
in the gauge theory. Note that this correspondence also allows us to
interpret the parameter $M$ in the gravity solution: it is the mass
one would obtain in Newtonian gravity. As in the
Schwarzschild example of ref.~\cite{Monteiro:2014cda}, electric charge
in the gauge theory then maps to mass in the gravity theory, provided
a suitable definition is used for the latter.

\subsubsection{Orbits and perihelion precession}\label{sec:orbits}

\begin{figure}[!b]
	\includegraphics[scale=0.33]{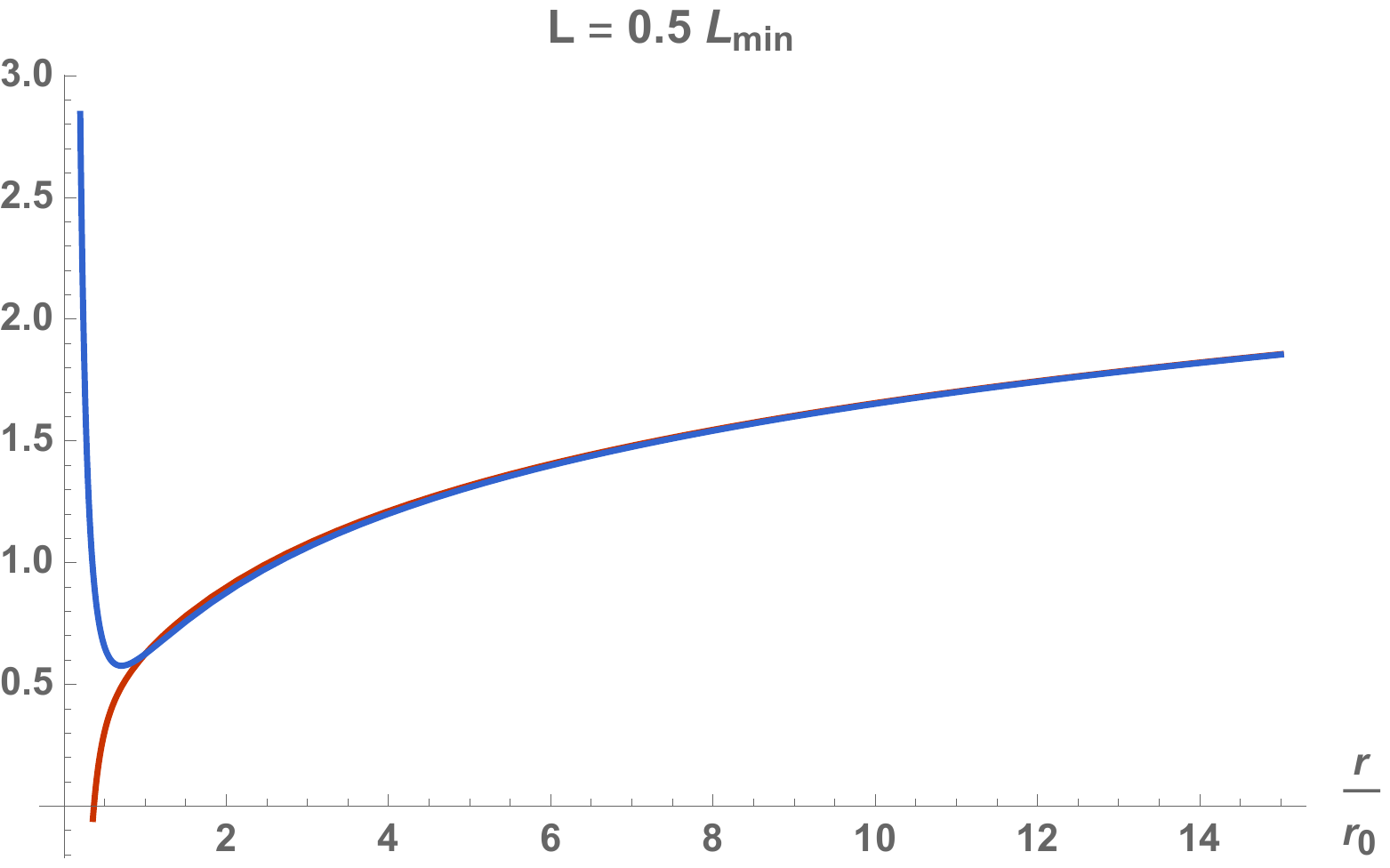}	\includegraphics[scale=0.33]{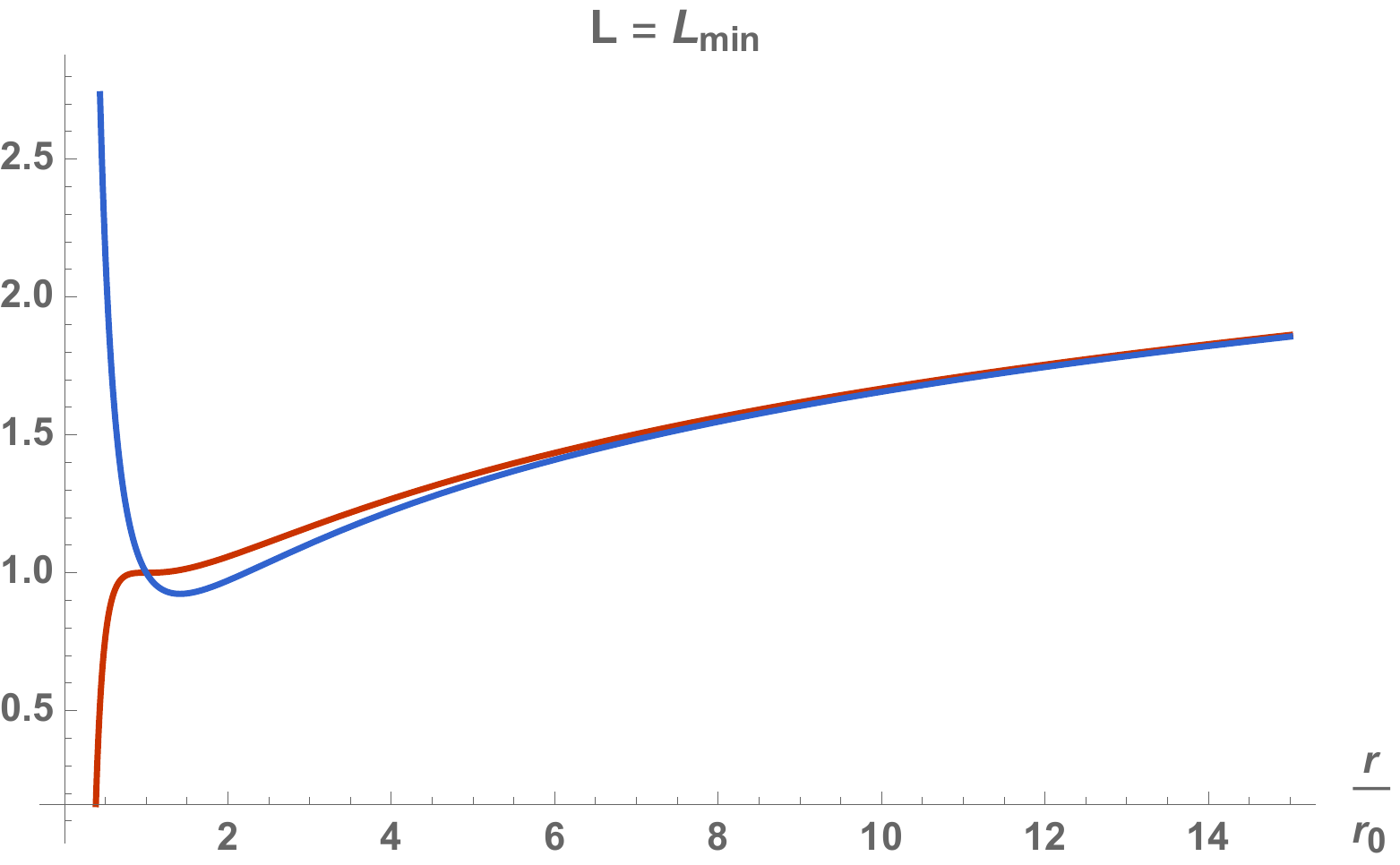}	\includegraphics[scale=0.33]{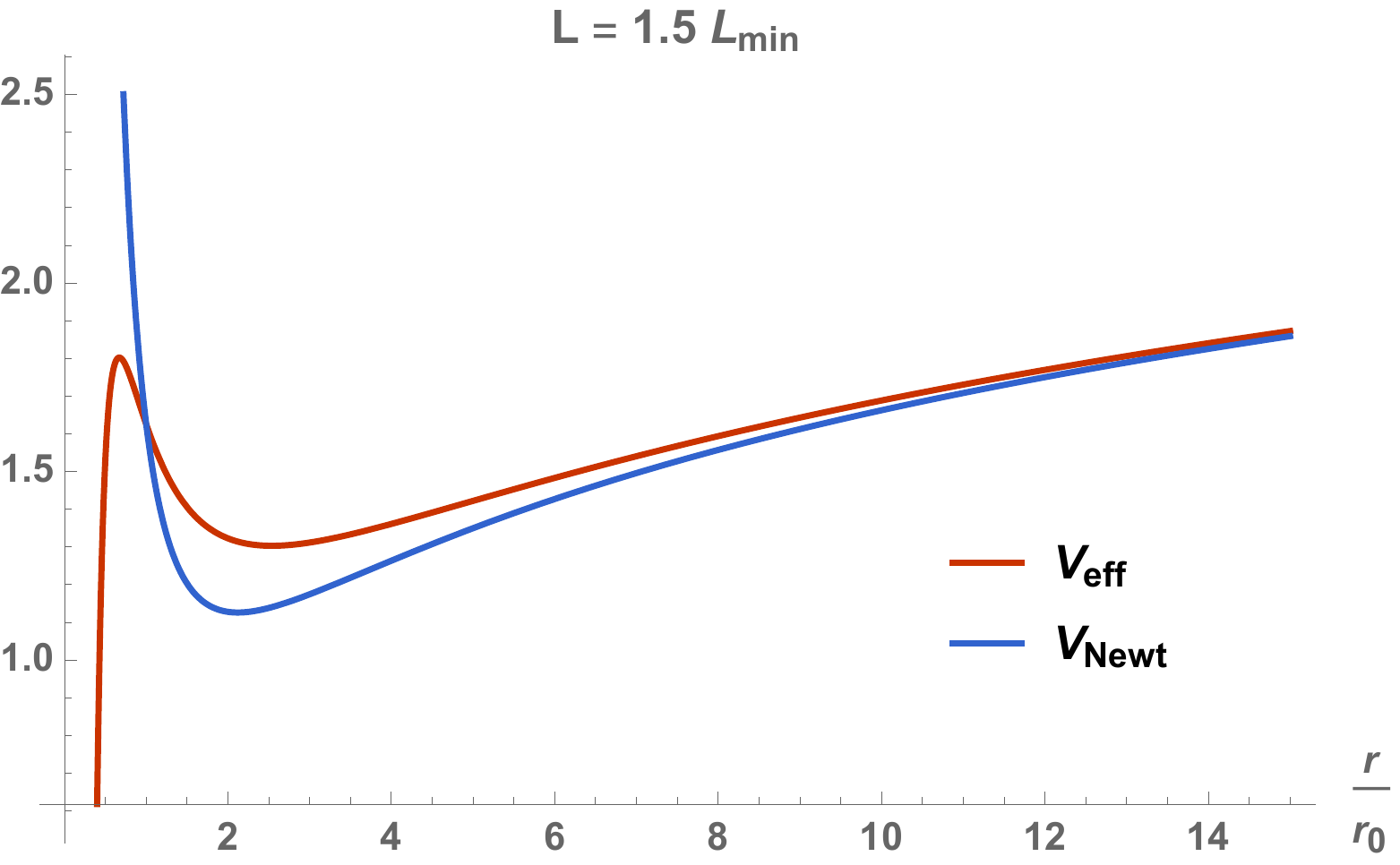}
	\caption{Effective GR and Newtonian potentials for $GM=1/2$  and  different values of the angular momentum $L$. We can clearly observe that  for $L<L_\text{min}$ the GR potential has no maximum, at $L=L_\text{min}$ it develops an inflection point, and for $L>L_\text{min}$ a  maximum  occurs  and we can have stable orbits. On the  other hand, the shape  of the Newtonian potential stays  unchanged  and always allows stable orbits.} \label{veff}
\end{figure}

To gain yet more physical insight into our solution, it is instructive
to examine the behaviour of geodesics corresponding to the motion of
(massive) test particles. To this end, notice that the metric of
Eq.\eqref{metric} can be written in ``Schwarzschild'' coordinates by
performing the coordinate transformation 
		\begin{displaymath}
			\ud t 
				= \ud T +\left((1+2GM\log{r})^{-1} - 1\right) \ud r \ , 
		\end{displaymath}
in which coordinates the metric becomes
		\begin{equation}
			\ud s ^2 
				= -\left(1+2GM \log{r}\right)\ud T^2 
					+\left(1+2GM \log{r}\right)^{-1}\ud r^2
					+r^2\ud\phi^2 . \label{SchwCoord}
		\end{equation}
In these coordinates the geodesic equation for
timelike observers can be written as
		\begin{eqn}\label{geodesic}
			\frac{1}{2}E^2
				&=\frac{1}{2}\left(\frac{\ud r}{\ud T}\right)^2 
					+ V_\text{eff},
			&
				V_\text{eff}
					&= \frac{1}{2}\left(\frac{L^2}{r^2}+1\right)
						\left(1+2GM\log{r/r_0}\right),
		\end{eqn}
where we have reintroduced the scale $r_0$ i.e. a constant radius
whose value is set by boundary conditions. In the expression above, we
have used the conserved quantities related to the timelike $\xi_t$ and
angular $\xi_\theta$ Killing vectors:
\begin{equation}
E=-g_{\mu\nu}\xi_t^\mu u^\mu \ , \quad  L=g_{\mu\nu}\xi_\theta^\mu u^\mu \,
\end{equation}
where $u^\mu$ is a unit timelike vector tangent to the
geodesic. Physically, $E$ and $L$ represent the energy and angular
momentum (relative to the origin) of the test particle. In
Eq.~(\ref{geodesic}), we have introduced the so-called {\it effective
  potential} $V_\text{eff}$. In the Newtonian limit, a similar
geodesic equation is obtained, but where the effective potential is
replaced by
\begin{equation}
V_\text{Newt}=\frac{1}{2}+\frac{L^2}{2 r^2}+GM\log{r/r_0},
\label{VNewt}
\end{equation}
which we obtain from Eq. (\ref{geodesic}) by neglecting the $L^2 GM$ term.
As is suggested by the form of Eq.~(\ref{geodesic}), values of the
parameters $(E,r_0,L)$ exist such that timelike geodesics form closed
orbits. We may find the relevant conditions by examining the form of
the effective potential, which is plotted for various values of $L$ in
figure~\ref{veff}. We see that the Newtonian potential always has an
infinite barrier at short distances, whilst the GR result turns over
at a finite distance. Furthermore (and in marked contrast to the case
of the Schwarszchild black hole in four spacetime dimensions), a
timelike observer can never escape to infinity, due to the fact that
the effective potential is logarithmically divergent there. This can
be traced back to the logarithmic behaviour of the gauge theory
propagator in two spatial dimensions, which enters the effective
potential via the double copy.

\begin{figure}[!h]
	\includegraphics[scale=0.38]{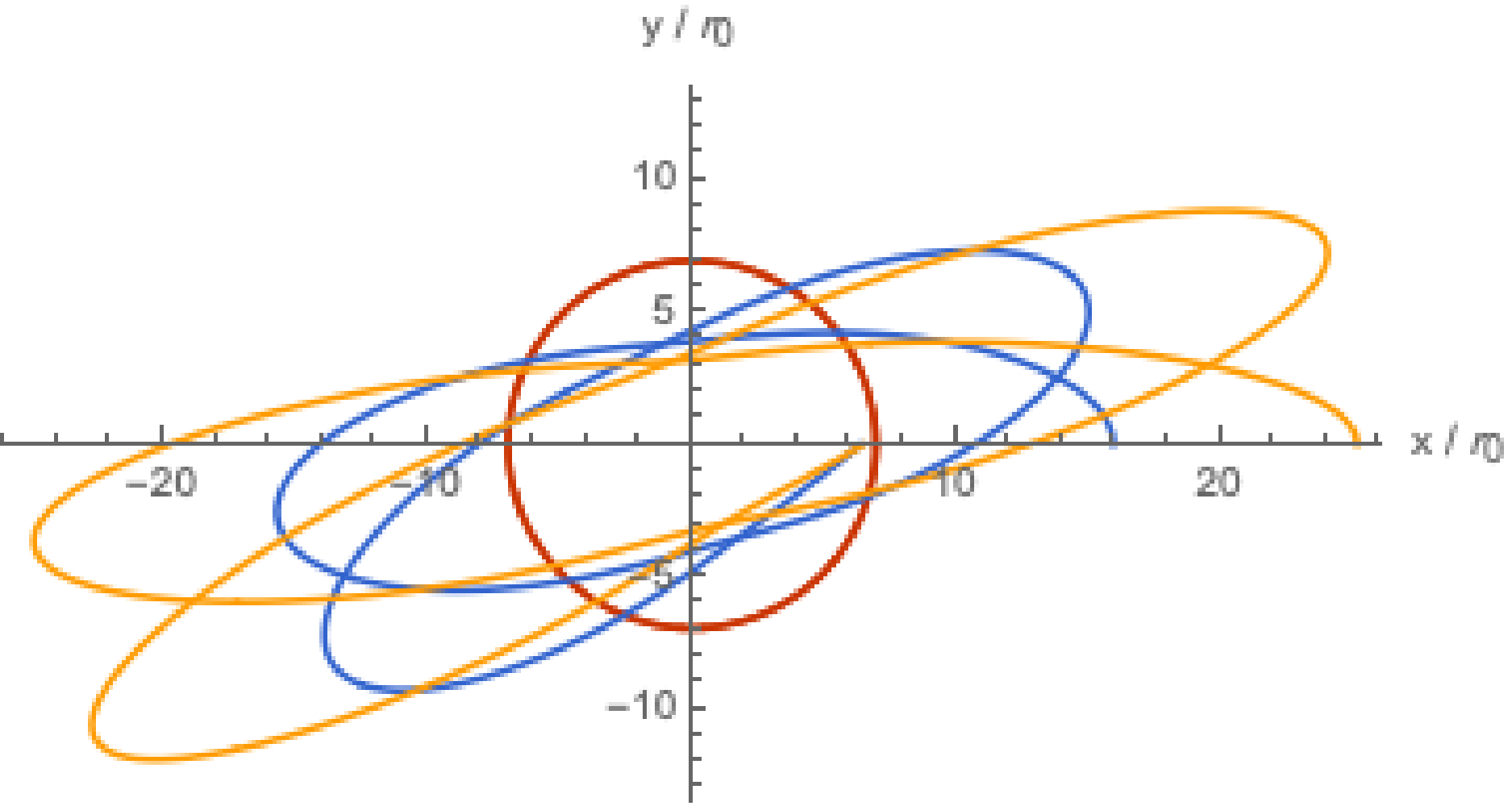} 	\includegraphics[scale=0.38]{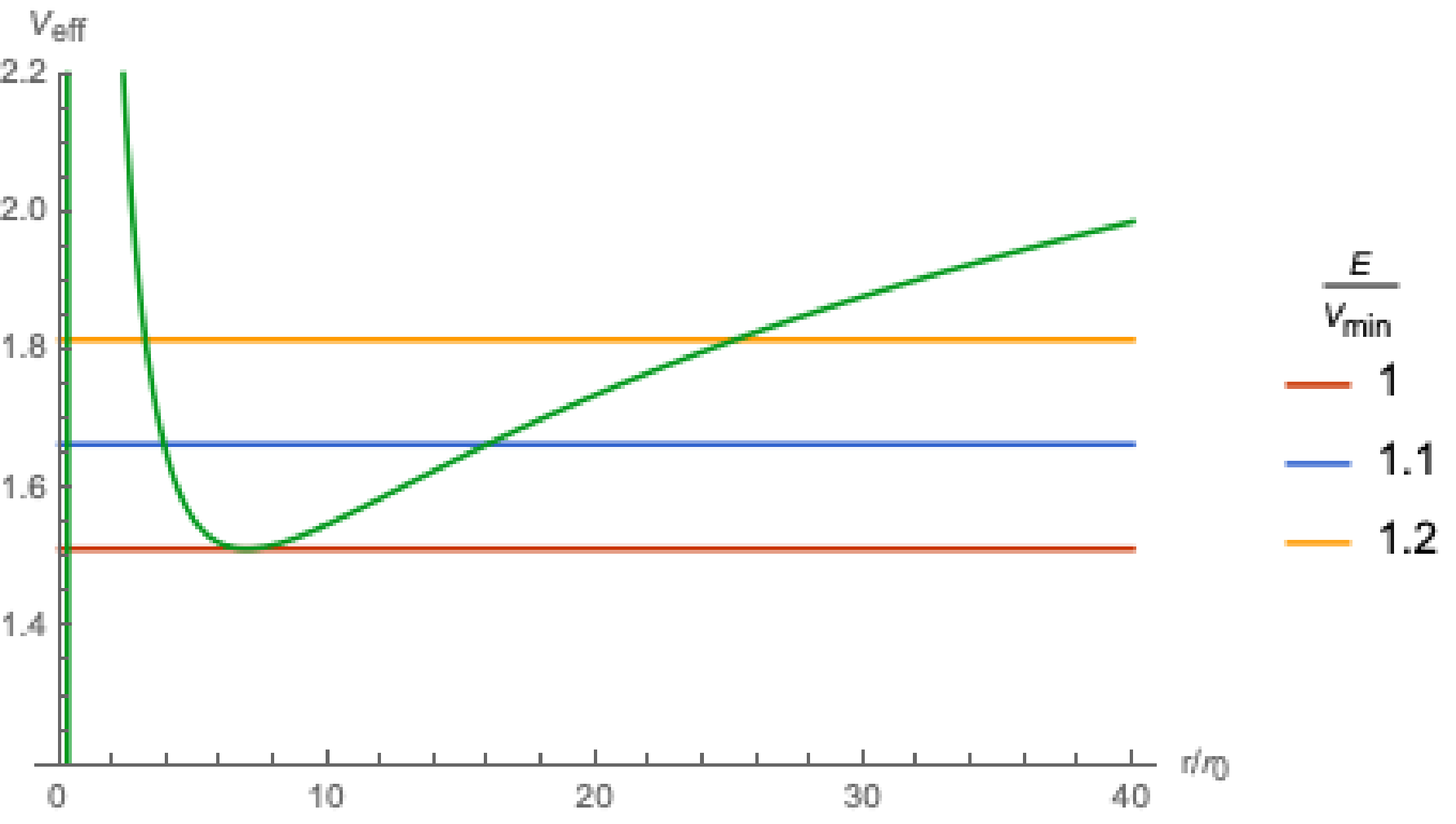}
	\caption{Timelike geodesics obtained for the GR effective
          potential, where $GM=0.4$ and
          $L/r_0=3$.} \label{normalorbits}
\end{figure}

\begin{figure}[!h]
	\includegraphics[scale=0.38]{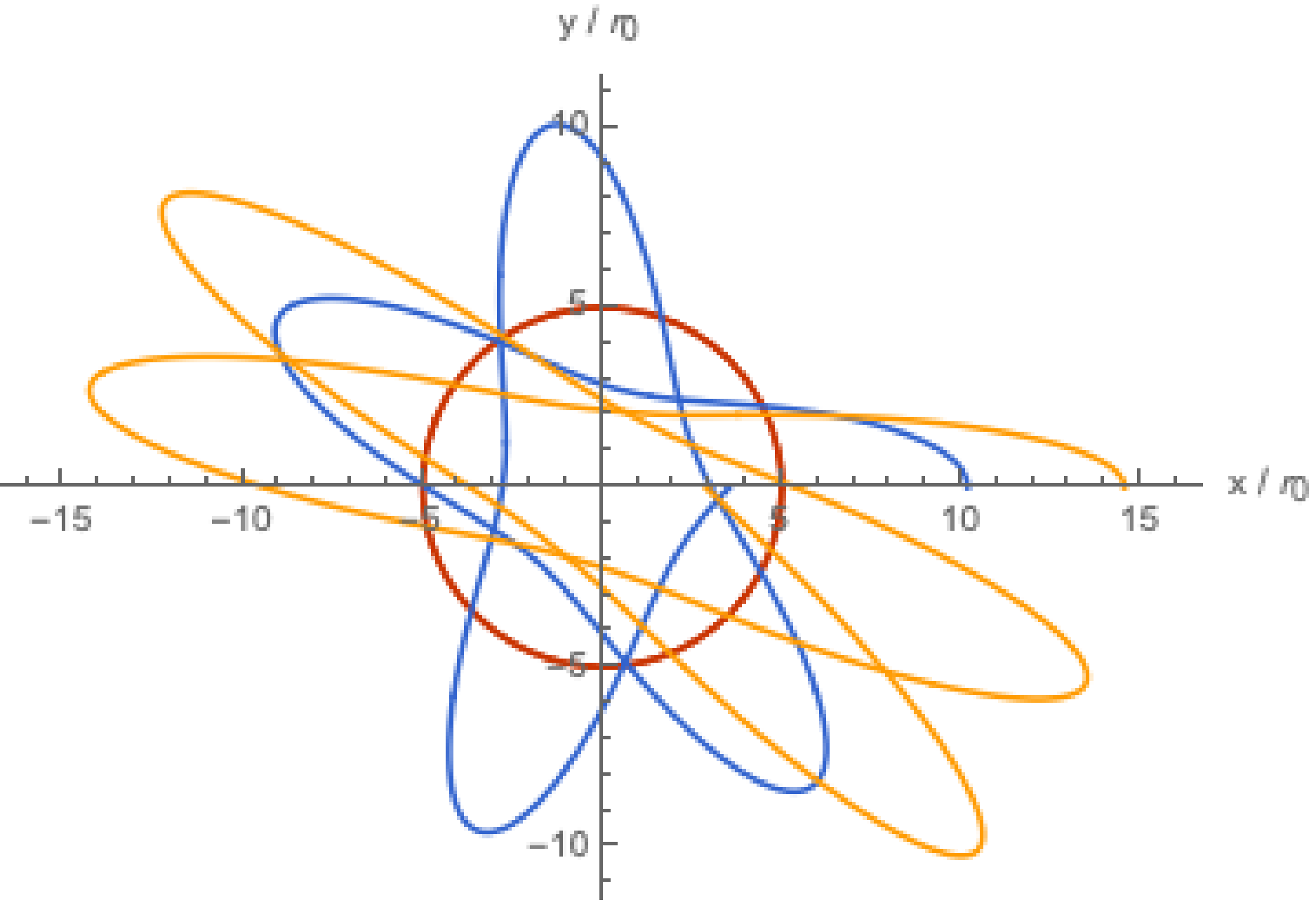} 	\includegraphics[scale=0.38]{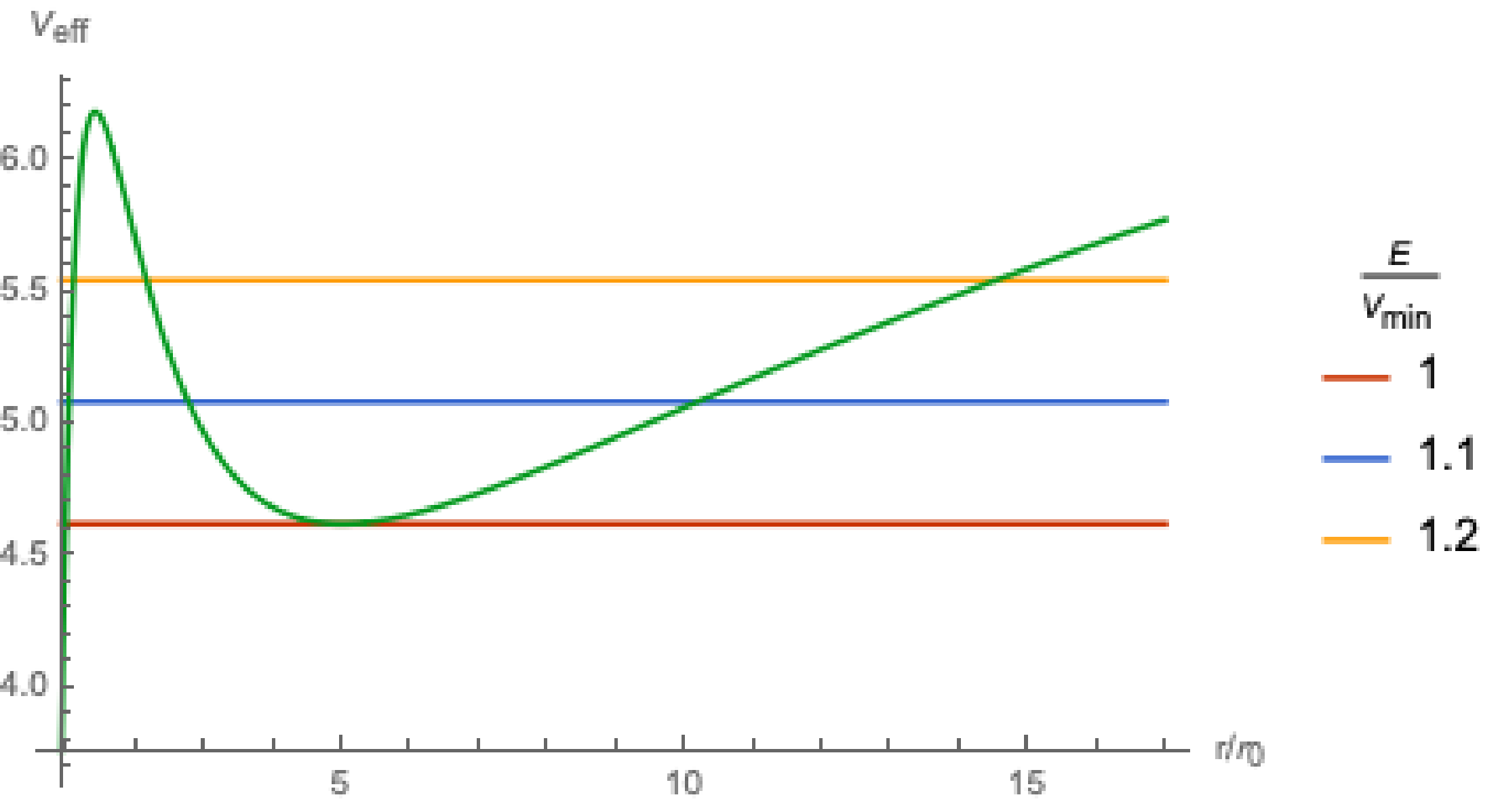}
	\caption{Timelike geodesics obtained for the GR effective
          potential, where $GM=1.8$ and $L/r_0=3$.} \label{starorbits}
\end{figure}

Examining the GR effective potential in more detail, we find that it
has a maximum as long as
\begin{equation}
L>L_\text{min}\equiv r_0 \, e^{1-\frac{1}{2GM}}  \ .
\end{equation}
We will therefore impose this constraint in order to obtain stable
orbits. In the same spirit, we will restrict the energy range such that
\begin{displaymath}
V_\text{eff}^\text{min}<E<V_\text{eff}^\text{max}, 
\end{displaymath}
where the maximum of the potential can be approximated for $L\gg r_0$
as follows:
\begin{equation}
V_\text{eff}^\text{max}\simeq \frac{1}{2}G M \left(\frac{L}{r_0}\right)^2 e^{\frac{1}{ G M}-1}\ .
\end{equation}
Meanwhile, we can only find an analytic expression for the minimum of
the potential in the Newtonian limit where $GML^2/r^2\ll 1$, yielding
\begin{equation}
V_\text{eff}^\text{min}\simeq  \frac{1}{2}+ \frac{1}{2}G M + G M \log \left(\frac{L}{\sqrt{GM} r_0}\right)\  .
\end{equation}
In order to visualise the orbits we must solve for $r(\theta)$. By using
the fact that $\dot\theta=L/r^2$, where the dot indicates a derivative
with respect to $T$ and $\theta$ is the azimuthal angle, we get
\begin{equation}
\left(\frac{\ud r}{\ud \theta}\right)^2-\frac{E^2\,  r^4}{L^2}+\frac{r^4}{L^2}+\frac{r^4 \,2GM\, \log{r}}{L^2}+r^2+ r^2\,2GM\, \log{r}=0 \ ,
\end{equation}
To further analyze this equation, it is convenient to change variables to the dimensionless
quantity $x=L/GMr$, and apply a derivative with respect to $\theta$
(denoted by a prime in what follows):
\begin{equation}
\label{GRorb}
G M x''+ G^2 M^2 \left(2x \log \left(\frac{L}{G M x}\right)- x\right)+G M x-\frac{1}{x}=0 \ .
\end{equation}
Taking the limit where $GM\ll 1$ we obtain
\begin{equation}
\label{Newtorb}
x''+x-\frac{1}{ G M x}=0 \ ,
\end{equation}
which indeed corresponds to the expected Newtonian limit given by the
potential in Eq.\eqref{NewtPot}.  In both the GR and Newtonian cases,
we may solve the radial equation numerically, and plot the orbits in
the $(r,\theta)$ plane. Considering first the GR case, we find
qualitatively different solutions, according to the value of the
quantity
\begin{equation}
\alpha=\frac{L^2}{r_0^2}  GM.
\label{alphadef}
\end{equation}
For small values of $\alpha$, one obtains a circular orbit if
$E=V_\text{min}$, and a precessing elliptical orbit otherwise, as
shown in Fig.~\ref{normalorbits}. Alternatively, for sufficiently high
values of $\alpha$ we find that the orbits have an interesting
``starfish'' shape, owing to the increasingly repulsive nature of the
potential barrier at small radial distances. This is shown in
Fig.~\ref{starorbits}. In both figures, we also show the form of the
effective potential in each case, with colour-coded values for the
energy $E$.

\begin{figure}[!h]
  	\includegraphics[scale=.6]{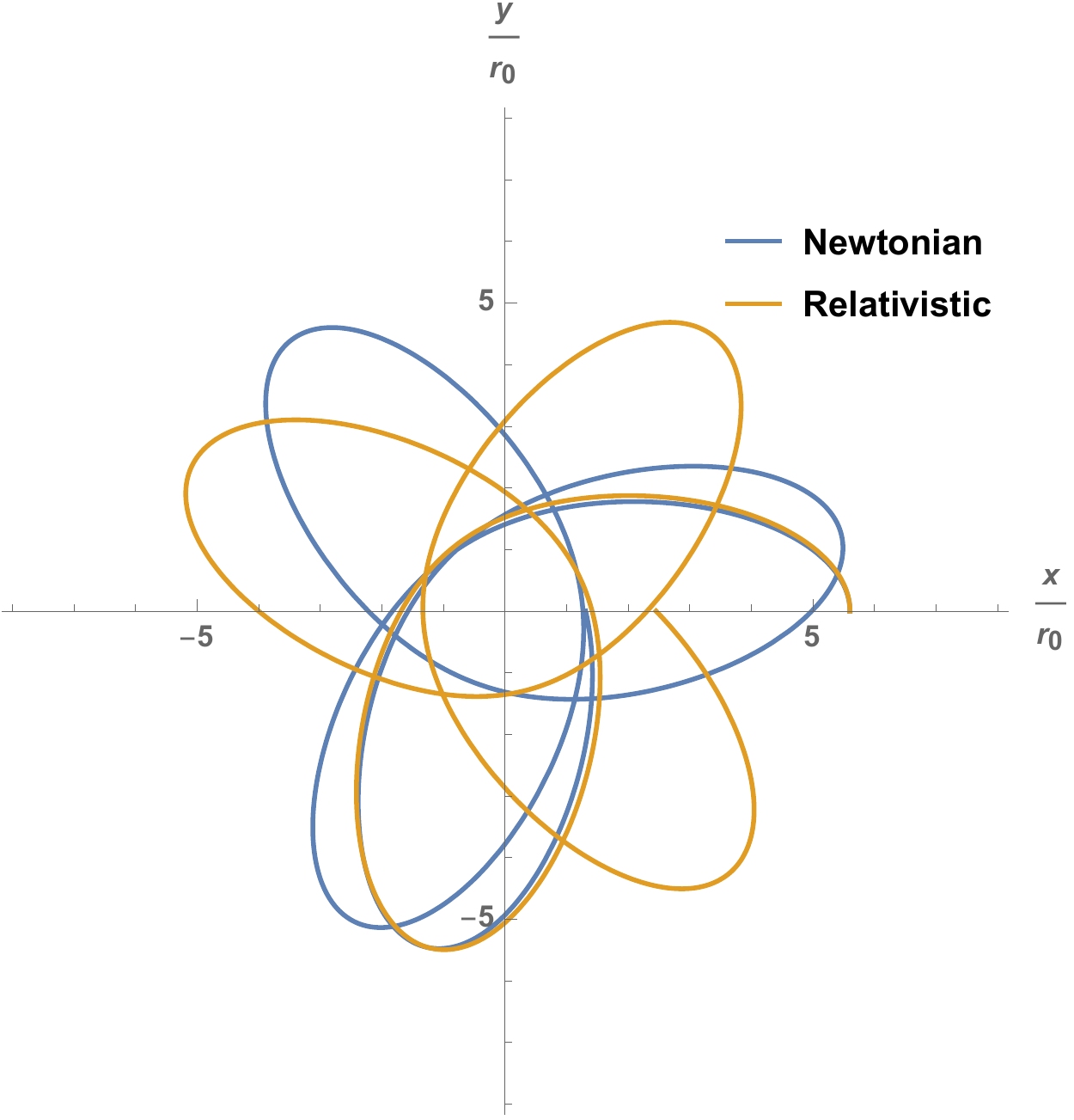}
 	 \caption{Bounded orbits obtained via solving Eqns \eqref{Newtorb} (displayed as the blue line) and \eqref{GRorb} (displayed as the orange line) with $GM=.2$, $L/r_0=1$, and $E=1.1 V_{min}$.}
	 \label{OrbitDiff}
  \end{figure}

The phenomenon of precession is common to the case of the
Schwarzschild metric in $d=4$, as indeed it must be: {\it Bertrand's
  theorem} states that the only central forces that give rise to
stable non-precessing orbits are an inverse square law, or a linear
radial dependence characteristic of a harmonic oscillator. The GR
effective potential in the present case is neither, and thus we expect
to see precession. A further consequence of this theorem, however, is
that we also expect to see precession in the Newtonian case, unlike
the case of Newtonian gravity in $d=4$ (i.e. the latter obeys an
inverse square force law). This is shown in Fig. \ref{OrbitDiff}, where we also
show an orbit obtained from the full GR potential for identical
initial conditions. We see that switching from the Newtonian theory to
the full GR case enhances the precession, analogous to the
Schwarzschild case. 

\section{Conclusion}
\label{sec:conclude}

In this paper, we have examined the double copy of exact classical
solutions in $(2+1)$ dimensions. This procedure has been defined for
any time-dependent Kerr-Schild solution in
gravity~\cite{Monteiro:2014cda} and studied in a number of non-trivial
examples~\cite{Luna:2015paa,Luna:2016due,Berman:2018hwd}, but poses
unique conceptual puzzles in three spacetime dimensions. In
particular, three dimensional gravity in the form of pure General
Relativity is known to have no propagating degrees of freedom, and
also lacks a sensible Newtonian limit. It is then not clear how
physical degrees of freedom in the gauge theory can be matched to the
gravity theory in a consistent manner. 

We first reviewed an existing
analysis~\cite{Carrillo-Gonzalez:2017iyj} of a pure gravity-like
solution, namely the BTZ black hole~\cite{Banados:1992gq}, whose
Kerr-Schild form was first presented in ref.~\cite{Kim:1998iw}. Upon
taking the single copy, the only surviving remnant in the gauge theory
is a constant charge density filling all space, which corresponds with
the cosmological constant (a constant energy density) in the gravity
theory, in accordance with the four-dimensional results of
ref.~\cite{Luna:2015paa}. 

Next, we considered a pointlike electric charge in (2+1)-dimensional
gauge theory, for which the Kerr-Schild double copy is
straightforward. This turned out to yield a non-vacuum solution of the
Einstein equations, in accordance with general observations we made
that localised sources in gauge theory cannot lead to vacuum solutions
under the Kerr-Schild double copy. The source term on the gravity side
could be interpreted as a non-zero dilaton profile, which is
consistent with the fact that the double copy of pure gauge theory is
{\it not} General Relativity, but rather GR coupled to a dilaton and
two-form. This is itself highly novel, in that our example is the
first case of the Kerr-Schild double copy in which the dilaton is
non-zero on the gravity side of the correspondence. Furthermore, it
resolves the issue of how to map the physical degrees of freedom in
the gauge theory: with no propagating graviton, the dilaton has to
turn on to do gravity's job. This in turn leads to a well-behaved
Newtonian-like limit, such that charge in the gauge theory maps to mass on
the gravity side, where the mass parameter can be interpreted by
consistency with the correct gravitational potential in the Newtonian
limit. This construction is reminiscent of previous attempts to obtain
sensible Newtonian limits in various scalar-tensor
theories~\cite{Verbin:1994ks,Barrow_1986}. To physically illustrate
our analysis, we examined timelike geodesics in the dilaton-gravity
system obtained by copying the point charge. These consist (for
appropriate initial conditions) of stable, precessing orbits. The
Newtonian orbits also precess in $d=3$, but the effect in the GR case
is exacerbated, analogous to the case of $d=4$.

An outstanding property of this dilaton-gravity solution obtained as the double copy is the existence of an event horizon at
\begin{equation}
r_h=r_0 \, e^{-\frac{1}{2GM}} \ ,
\end{equation} 
which can be observed from the metric in Schwarzschild coordinates in Eq.\eqref{SchwCoord}. This  implies that the solution corresponds to a  phantom black hole (since the dilaton in this generalized gauge has the incorrect sign for the  kinetic term). In the present case, the no-hair theorem  \cite{Bekenstein:1972ny,Hawking:1972qk,Bekenstein:1995un} is avoided since the scalar field is a ghost and it does not  inherit the symmetries of the spacetime, that is, $\partial_\theta \varphi\neq0$. Phantom black holes and their thermodynamic properties have previously been studied, for example in \cite{Gibbons:1996pd,Rodrigues:2012tw}. It would be interesting to study the thermodynamic properties of this 3-dimensional phantom black hole, and we leave this as future work. 

The study of the double copy for classical solutions is still in its
infancy, and useful insights can be gained by examining highly unusual
cases, even if these are not directly physical by themselves. We hope
that our results prove useful in this regard, and stimulate further
work in this area.

\section*{Acknowledgments}

CDW was supported by the Science and Technology Facilities Council
(STFC) Consolidated Grant ST/P000754/1 \textit{``String theory, gauge
  theory \& duality''}, and by the European Union's Horizon 2020
research and innovation programme under the Marie Sk\l{}odowska-Curie
grant agreement No.~764850 {\it
  ``\href{https://sagex.ph.qmul.ac.uk}{SAGEX}''}. The research of BM,
KR, and SW was supported in part by DOE grant DE-FG02-85ER40237. The
work of M.C.G. was supported in part by US Department of Energy (HEP)
Award DE-SC0013528.

\appendix
\section{No Fluid Interpretation of the Double Copy Stress-Energy Tensor}  \label{nofluid}

In section~\ref{sec:chargeDC}, we have seen that the stress-energy
tensor generated by double-copying a gauge theory point charge
(Eq.~(\ref{stresstensor})) in (2+1) dimensions can be associated with
a dilaton profile.  In this appendix, we ask whether it is possible to
instead associate this with a viscous fluid. For the latter, the
stress tensor has the generic form
\begin{equation}
T_{\mu\nu}=\rho\, u_\mu u_\nu + P\, \gamma_{\mu\nu}  - \zeta \theta \gamma_{\mu\nu}+2 \eta\sigma_{\mu\nu}+\Pi^\rho\,\gamma_{\rho(\mu}u_{\nu)} \ .  \label{st}
\end{equation}
where $\rho$ is the energy density, $P$ the pressure, $\eta$ the shear
viscosity, $\zeta$ the bulk viscosity,
$\theta\equiv\gamma^{\nu\mu}\nabla_\nu u_\mu$ the expansion, and
$\Pi^\rho$ the momentum density. The metric of the surface
perpendicular to the four velocity $u^\mu$ is
\begin{equation}
\gamma_{\mu\nu} = g_{\mu\nu} + u_{\mu}u_{\nu} \ ,
\end{equation}
where the four velocity is normalized such that $u\cdot u=-1$. Lastly,
the shear tensor $\sigma_{\mu\nu}$ is
\begin{equation}
\sigma_{\mu\nu}\equiv\frac{1}{2}\left(\gamma^\rho_\mu\nabla_\rho u_\nu+\gamma^\rho_\nu\nabla_\rho u_\mu\right)-\frac{1}{2}\gamma_{\mu\nu} \theta \ .
\end{equation}
In vacuum (2+1) dimensional gravity there is no shear, but the
situation can change in the presence of matter.

We will now assume that the stress tensor for the double copy of the
static charge can be written as in Eq.\eqref{st}. With this
assumption, the fluid properties can be extracted from the stress
tensor as
\begin{align}
&\rho=T_{\mu\nu}u^\mu u^\nu \ , \\
&2(P- \zeta \theta)=T_{\mu\nu}\gamma^\mu_\rho\gamma^{\nu\rho} \ , \\
&\Pi^\rho=T_{\mu\nu}\gamma^{\rho\nu} u^\mu \ .
\end{align}
Working in Kerr-Schild coordinates, that is, with the metric  given by Eq.\eqref{metric}, we  can see that  the four velocity is  given by
\begin{equation}
u^\mu=\frac{1}{\sqrt{1+4 G M  \log(r)}}\left(1,0,0\right)  \ , \label{vel}
\end{equation}
and thus
\begin{equation}
\rho=-\frac{M}{ 4\pi r^2} \ , \quad P- \zeta \theta=P=0 \ , \quad \Pi^\rho=(0,0,0)\  , \quad \sigma_{\mu\nu}= \mathbf{0}   \ .  \label{propfluid}
\end{equation}
From these, we see that the covariant derivative of the four-velocity can be written as
\begin{equation}
\nabla_\mu u_\nu= -\dot{u}_\mu u_\nu \ .
\end{equation}
This describes changes in the relative position of the worldlines of
two neighboring timelike observers. Note that the shear, vorticity,
and expansion are zero. Thus, the shape of the cross-sectional area
orthogonal to the timelike geodesic flow is not deformed as we move
along the geodesics. This behavior is specific for the congruence
constructed above, but in general it could be possible to find cases
where the area orthogonal to the flow is deformed.

Using the above result, we can find a discrepancy in the assumption that the stress tensor of Eq.\eqref{stresstensor} is the stress tensor of a viscous fluid with four-velocity given by Eq.\eqref{vel}. While for a fluid with the properties in Eq.\eqref{propfluid} we  have
\begin{equation*}
T^\text{fluid}_{\rho\lambda}\gamma^\rho_\mu \gamma^\lambda_\nu=P\, \gamma_{\mu\nu}  - \zeta \theta \gamma_{\mu\nu}+2 \eta\sigma_{\mu\nu}= \mathbf{0} \ ,
\end{equation*}
for the stress tensor in Eq.\eqref{stresstensor}  we have
\begin{equation*}
\Sigma_{\mu\nu}\equiv T_{\rho\lambda}\gamma^\rho_\mu \gamma^\lambda_\nu= \frac{M}{4  \pi } \left(
\begin{array}{ccc}
0 & 0 & 0 \\
0 & \frac{1 }{ (1+4GM  \log(r)) r^2} & 0 \\
0 & 0 & -1 \\
\end{array}
\right) \ .
\end{equation*}
Thus, the stress tensor of  the double copy can be written as 
\begin{equation}
T_{\mu\nu}=\rho\, u_\mu u_\nu +\Sigma_{\mu\nu} \ .  
\end{equation}
Since $\Sigma_{\mu\nu}$ cannot be interpreted as a property of a
fluid, this implies that the stress tensor of Eq.\eqref{stresstensor}
does not correspond to a viscous fluid.  Note that we could identify
the stress tensor in Eq.\eqref{stresstensor} with that of a
``Euclidean'' perfect fluid if we consider a spacelike
four-velocity. This is easily seen by considering
$u^\mu\propto\partial^\mu \varphi$ where $\partial^\mu \varphi$ is
given by Eq.\eqref{derphi}.

\bibliographystyle{apsrev4-1}
\bibliography{refs}

\end{document}